\documentclass[%
 reprint,
 amsmath,amssymb,
 aps,
]{revtex4-1}
\usepackage[utf8]{inputenc}
\usepackage{graphicx}
\usepackage{slashbox}
\usepackage{sidecap}
\usepackage{cancel}
\usepackage{amsthm,amsmath}
\usepackage{framed}
\usepackage[font={footnotesize,it}]{caption}
\usepackage{color}
\usepackage{marvosym}
\usepackage{amssymb}
\theoremstyle{definition} 

\newcommand{\vast}{\bBigg@{4}}
\newcommand{\Vast}{\bBigg@{5}}

\newcommand{\eq}[1]{\,\begin{equation}
                   #1 
                   \end{equation}
}

\newcommand{\morabba}[1]{\,\begin{flushright}
 \Rectsteel \\
\end{flushright}}

\newcommand{\eqq}[2]{\,\begin{equation} \label{#2}
                   #1 
                   \end{equation}
}

\usepackage{setspace}

\newcommand{\al}[1]{\,\begin{align}
                   #1 
                   \end{align}
}
\newcommand{\all}[2]{\,\begin{align}
                   #1 
                    \label{#2}
                   \end{align}
}

\usepackage[top=24mm, bottom=25mm, left=22mm, right=20mm]{geometry}

\makeatletter

\newcommand{\CC}[2]{\, B^{#1}_{#2} 
}
\makeatother

\newcommand{\CCC}[2]{\, C^{#1}_{#2} 
}

\begin{document}
\preprint{APS/123-QED}

\title{ Degree Correlation in Scale-Free Graphs}

\author{Babak Fotouhi and Michael G. Rabbat \\
Department of Electrical and Computer Engineering\\
McGill University, Montr\'eal, Qu\'ebec, Canada\\
Email:\texttt{ babak.fotouhi@mail.mcgill.ca, michael.rabbat@mcgill.ca}\\}


\begin{abstract}
We obtain closed form expressions for the expected conditional degree distribution and the joint degree distribution of the linear preferential attachment model  for  network growth in the steady state.  We consider the multiple-destination preferential attachment growth model, where incoming nodes at each timestep attach to~$\beta$  existing nodes, selected by degree-proportional probabilities. By the conditional degree distribution $p(\ell| k)$, we mean the degree distribution of  nodes that are  connected  to a node of degree $k$. By the joint degree distribution $p(k,\ell)$, we mean the proportion of links that connect nodes of degrees~$k$ and~$\ell$. In addition to this growth model, we consider the shifted-linear preferential growth and solve for the same quantities, as well as a closed form expression for its steady-state degree distribution.

\end{abstract}
\maketitle

\section{Introduction}

The degree distribution of a network, denoted by~$n_k$ in this paper, is the probability that  a randomly-chosen node will have~$k$ links. An array of  networks  have been observed to exhibit a power-law degree distribution, e.g.,  the World Wide Web~\cite{WWW_SF_1,WWW_SF_2}
, biological networks~\cite{biol_1,biol_2}, scholarly citation networks~\cite{collab_1, collab_2} and collaboration networks of film  actors~\cite{actors_1, actors_2}. 

Among the models of network formation proposed to emulate the scale-free property (the power-law tail) is the linear preferential attachment scheme, introduced in~\cite{barabasi_SF}. In this model, the graph is constructed by successive addition of nodes. The initial graph is taken to be a complete graph, but in the long-time limit, the initial conditions have  no effect. Each node that is added  chooses~${\beta \geq 1}$ existing nodes and attaches to them. We call these~$\beta$~nodes the \emph{parents} of the new node. The probability of existing nodes being  picked by the new node is proportional to their degrees. If we denote the set of all nodes in the network at time~$t$ by~$V(t)$, then a node with degree~$k$ receives a link from the new node with the following probability
\eq{
P_t(k)=\frac{k}{\displaystyle \sum_{i: i \in V(t)} k_i}
.}

In addition to the degree distribution,  the \emph{conditional degree distribution}~$p(\ell|k)$ is an important structural quantity. It is defined as the probability that a randomly chosen neighbor of a node of degree~$k$  will have degree~$\ell$. This quantity is invoked for example in studying percolation processes over networks~\cite{xulvi, hooyberghs, schwartz}, resilience of complex networks against random attacks~\cite{deng, gallos}, and spread of epidemic diseases over networks~\cite{boguna_epid, moreno_epid}. 

Another structural measure of networks is the \emph{joint degree distribution}~${p(k,\ell)}$, the fraction of links whose incident nodes have degrees~$k$~and~$\ell$. Note that   the joint degree distribution is  symmetric by definition; that is,  ~${p(k,\ell)=p(\ell,k)}$.  

In this paper, we find closed form expressions for~$p(\ell|k)$ and~${p(k,\ell)}$ in the limit as~$t \rightarrow \infty$. 
Let~$\CC{x}{y}$ denote the binomial coefficient~$ \binom{x}{y}$.
For the expected conditional degree distribution, we show that
\al{
p(\ell | k)= \displaystyle \frac{\beta(k+2)}{k \ell (\ell+1)}
-\displaystyle \frac{\beta}{k \ell} \CC{2\beta+2}{\beta+1}
\frac{\CC{k+\ell-2\beta}{\ell-\beta}}{\CC{k+\ell+2}{\ell}}
.}
For the joint degree distribution, our finding is
\al{
p(k,\ell)&= 
\displaystyle 
\frac{2 \beta(\beta+1) }{k(k+1) \ell (\ell+1)} \nonumber \\
&-\displaystyle \frac{2 \beta(\beta+1)}{k (k+1)\ell(\ell+1)} \CC{2\beta+2}{\beta+1}
\frac{\CC{k+\ell-2\beta}{\ell-\beta}}{\CC{k+\ell+2}{\ell+1}}
.}

Another class of preferential attachment  models are those with shifted-linear kernels~\cite{redner1}. The probability of attaching to an existing node of degree~$k$ is proportional to~${k+\theta}$, where~${\theta \geq 0}$ is a  constant.  
We obtain an expression for the conditional degree distribution. It is in the form of a finite sum
\al{
p(\ell|k)&=  
\displaystyle   \frac{\Gamma(k+1+\theta+\nu)  \Gamma(\ell+\theta)} 
{\Gamma(\beta+\theta) k \Gamma(k+\ell+2\theta+\nu+1)} \nonumber \\
& \times \Bigg[ 
\sum_{\mu=\beta+1}^k \displaystyle \frac{\Gamma(\mu+\beta+2\theta+\nu) }{\Gamma ( \mu+\theta+\nu )  } 
\CC{k+\ell-\beta-\mu}{\ell-\beta}
\nonumber \\
&+ 
\sum_{\mu=\beta+1}^{\ell} \displaystyle \frac{\Gamma(\mu+\beta+2\theta+\nu) }{\Gamma ( \mu+\theta+\nu )  } 
\CC{k+\ell-\beta-\mu}{k-\beta}
\Bigg] 
.}
In this paper we use the rate equation approach~\cite{redner2, drog_nk, redner1, redner3, redner4, redner5}, and solve for the desired quantities in the steady state.

The rest of the paper is organized as follows. First  in~Section~\ref{sec:notation} we introduce notation and the  basic  graph-theoretical approach undertaken to define and quantify~${p(\ell|k)}$ and~${p(k,\ell)}$, and then illustrate them through an example. In Section~\ref{sec1}  we focus on the preferential attachment model and solve the rate equations for~$n_{k \ell}$, which is defined as the fraction of nodes who have degree~$\ell$ and are connected to a parent node of degree~$k$. Then we use this quantity to obtain the desired conditional and joint degree distributions.   In Section~\ref{sec:shifted_sec}  we  focus on the shifted-linear preferential attachment and repeat those calculations after providing the motivation for studying this family of models.

\subsection{Notation, Terminology and Method} \label{sec:notation}

Throughout this article,  whenever a quantity is time-dependent, it will be explicitly mentioned. Otherwise, they represent the steady-state limit of the network growth process, i.e., as~${t \rightarrow \infty}$. The number of nodes of the network   is denoted by~$N$ and the fraction of nodes with degree~$k$ (the degree distribution) is denoted by~$n_k$. The number of nodes with degree~$k$ is denoted by~$N_k$ and we have ${N_k=N n_k}$.

The network growth process starts off from an initial graph, whose number of links is denoted by~$L(0)$. The following holds
\eq{
L(0)= \frac{1}{2} \sum_k k N_k(0)
.}

The initial network is assumed to be connected, and its characteristics play no role in the steady-state regime. ~$V(t)$ denotes the set of all nodes at time~$t$.

We denote the number of links  that  connect a pair of nodes with degrees~$k$ and~$\ell$ by~$L_{k\ell}$. By definition,~$L_{k\ell}$ is symmetric, that is, we have~${L_{k\ell}=L_{\ell k}}$.

Each node connects to~$\beta$ exiting nodes upon addition to the network. We call the new node the \emph{child} of those~$\beta$ nodes, and they are called the~\emph{parents} of the new node. By a \emph{parent-child pair of~${(k,\ell)}$}, we mean a  child of degree~$\ell$ who is connected to a parent of degree~$k$.

Let $N_{k\ell}$ denote the number of nodes with degree~$\ell$ who are connected to a parent node of degree~$k$. Note that the following holds: 
\all{
L_{k\ell}=N_{k\ell}+N_{\ell k}
.}{L_N_rel}
Also, we define
\al{
n_{k \ell} \stackrel{\text{def}}{=} \displaystyle \frac{N_{k\ell}}{N}
.}
As mentioned in the previous section, we denote the  conditional degree distribution by~$p(\ell|k)$ and the joint degree distribution by~${p(k,\ell)}$. The former is not symmetric with respect to~$k$~and~$\ell$, but the latter is. 

For any graph, to find~$p(\ell|k)$, one first has to count all the nodes with degree~$\ell$ who have a neighbor with degree~$k$ (taking degeneracy into account,  so that if  a node of degree~$\ell$ has, for example, two neighbors of degree~$k$, it must be counted twice). Dividing this number by the number of all nodes who have a neighbor of degree~$k$, gives~$p(\ell|k)$.

\begin{figure}[t]
  \centering
  \includegraphics[width=1 \columnwidth]{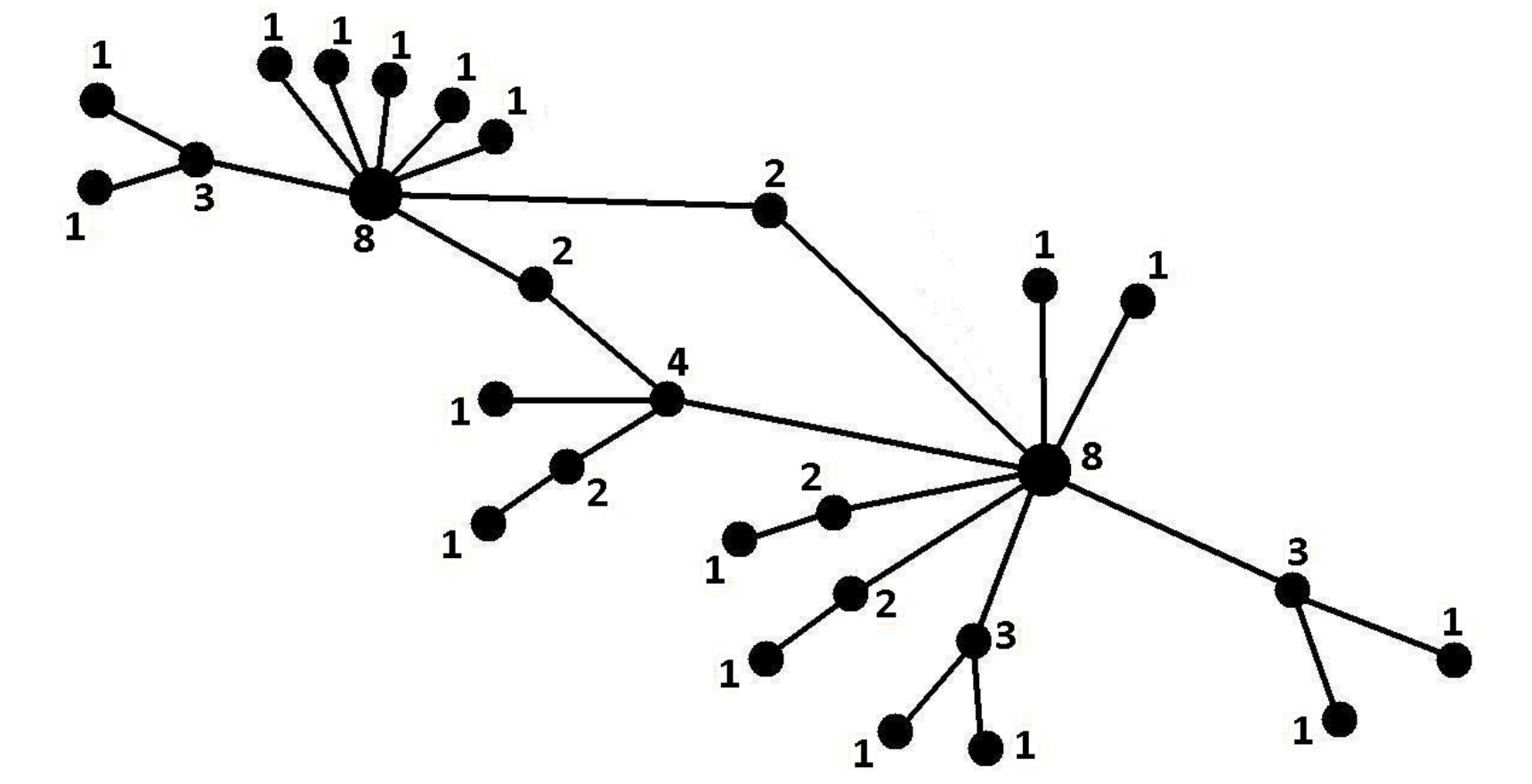}
  \caption[Figure ]%
  {An example graph used to illustrate the validity of~\eqref{pkl_def}. The number beside each node is its degree. }
\label{fig_graph}
\end{figure}

Take the graph shown in Fig.~\ref{fig_graph} as an example. Let us find~$p(3|8)$. There are two nodes with degree 8, so there are 16 nodes who are adjacent to a node of degree 8. So there is a set of 16 nodes, in which we search for nodes of degree
3. The left degree-8 node has one neighbor of degree 3, and the  right degree-8 node has two neighbors of degree 3. So we find
\eq{
p(3   |8)= \displaystyle \frac{3}{16}
.}
Note that the same result could be obtained as follows. First count the number of links who connect the nodes of degree 3 and 8. In this case, there exists  three such links. Then divide this by 16 (which is the total number of links that are incident to a node of degree $8$), that is, 
\eq{
p(3   |8)= \displaystyle \frac{L_{3,8}}{8 N_8}= \displaystyle \frac{3}{16}.
}

If we were looking for~$p(2|8)$, then the node in the middle who has degree 2, connecting the two nodes with degree 8, would be counted twice. That is, the left degree-8 node would give two neighbors of degree two, and the right one would give three such neighbors, and we would obtain
\eq{
p(2   |8)= \displaystyle \frac{5}{16}
.}
Same result is obtained using the link notation: 
\eq{
p(2   |8)= \displaystyle \frac{L_{2,8}}{8 N_8}= \displaystyle \frac{5}{16}
.}
So we find that the conditional degree distribution satisfies
\eqq{
p(\ell   |k)= \displaystyle \frac{L_{k,\ell}}{k N_k}.
}{pkl_def}

Similarly, for the joint degree distribution, we have
\al{
p(\ell,k)= \frac{L_{k,\ell}}{L} = \frac{p(\ell|k) k n_k}{\frac{1}{2} \sum_k k n_k}
.}

\subsection{Related Work: Rate Equation Approach, Uncorrelated Approximation}

In this paper, we employ the rate equation to capture the time evolution of~${N_{k\ell}(t)}$, and then we extract recursion relations for~$n_{k\ell}$  in the steady state and solve them. The rate equation is a potent approach in non-equilibrium statistical mechanics which was first  used to study network growth problems in~\cite{redner1, redner2, redner3, redner4, redner5, drog_nk}. 

In~\cite{redner1}, the special case of~${\beta=1}$ has been considered, and~$n_{k \ell}$ has been found. The result is
\all{
n_{k\ell}&= 
\displaystyle \frac{ 4(k-1)}{\ell(\ell+1)(k+\ell)(k+\ell+1)(k+\ell+2)} \nonumber \\
&+ \displaystyle \frac{  12(k-1) }{ \ell(k+\ell-1)(k+\ell)(k+\ell+1)(k+\ell+2)}
.}{eq_redner_nkl}
(Note the swap of $k,\ell$ due to  the  difference in notation).  We will show that our solution in the special case of~$\beta=1$  is  consistent with this  result. 

For Erdős–Rényi random graphs~\cite{random_1, random_2}, finding~${p(\ell |k)}$ is straightforward. In these graphs, nodes are uncorrelated and the degree distribution of the neighbors of each node is independent of that node's characteristics, including its degree. It coincides with the degree distribution of the node reached by following a randomly chosen link towards either end. A node with degree~$\ell$ has~$\ell$ links through which it can be picked. There are~$N_{\ell}$ nodes with degree~$\ell$ in the network, each with~$\ell$ of those links. So the probability of reaching a node of degree~$\ell$ under this mechanism is proportional to~$\ell N_{\ell}$. After normalization, we get
\eqq{
p_u(\ell |k)= \displaystyle \frac{\ell n_{\ell}}{ \displaystyle \overline{k}}
,}{pkl_uncorr_1}
where~$\bar{k}$ is the average degree of the whole network, and the subscript~$u$ refers to the uncorrelated approximation. This expression has been used as an~``uncorrelated approximation" to the scale-free networks, for theoretical analysis under the mean-field assumption in the thermodynamic limit. Examples include~\cite{Eguiluz_PRL,Barthelemy_PRL,Castellano_PRL,Castellano_PRL_2,boson,contact_process,cohen,dezso,wang,moreno_rumor_1,moreno,from_ieee_1,Xia,mizutaka,safety,langevin,Gallos_PRL,Cohen_book,Cohen_PRL_1,Cohen_PRL_2,Liu,Caccioli,Gallos_PRL_2,halvin_13,compete,pastor_PRL}. For a detailed list of examples, see~\footnote{Examples include Equation~(3) in~\cite{Eguiluz_PRL}, Equation~(3) in~\cite{Barthelemy_PRL}, Equation~(2) in~\cite{Castellano_PRL}, 
Equation~(3) in~\cite{Castellano_PRL_2}, Equation~(22) in~\cite{boson},  assumptions leading to Equation~(5) in~\cite{contact_process},  Equation~(6) in~\cite{cohen}, the tacit assumption behind Equation~(1) in~\cite{dezso},
Equation~(6) in~\cite{wang},  Equation~(9) in~\cite{moreno_rumor_1}, Equation~(10) in~\cite{moreno} which leads to  Equation~(21) and eventuates in Equations~(31) and~(42) for Scale-Free networks,  Equation~(2) in~\cite{from_ieee_1}, Equation~(3) in~\cite{Xia}, Equation~(10) in~\cite{mizutaka}, Equation~(8) in~\cite{safety}, and below Equation~(2) in~\cite{langevin}. In~\cite{Gallos_PRL}, absence of degree correlations  is implicitly assumed in the threshold value~$\kappa= \frac{\langle k^2 \rangle}{\overline{k}}$  for the existence of spanning trees. In~\cite{Cohen_book}, equation~(1.29) is obtained by applying the threshold value~$\kappa= \frac{\langle k^2 \rangle}{\overline{k}}$ for uncorrelated graphs to Scale-free networks. Same approximation is used in Equation~(9) of~\cite{Cohen_PRL_1}, Equation~(5) of~\cite{Cohen_PRL_2}, and Equation~(5) of~\cite{Liu}. In~\cite{Caccioli}  the uncorrelated approximation is used throughout the letter, mentioned at the end of the second paragraph of Section~(2).  The explicit form of the  conditional degree distribution  is required to  calculate the merit $E_b(k)$ introduced in Equation~(3) of~\cite{Gallos_PRL_2} in closed form, and also replace the asymptotic expression in  Equation~(1) of~\cite{Gallos_PRL_2}  used to solve the differential equation~(9) in the same paper.  In Section 2 of~\cite{halvin_13}, the branching factor that is introduced, which is then applied to scale-free networks, is obtained under the  assumption of uncorrelatedness. The  conditional degree distribution  of scale-free networks  is invoked in~\cite{compete}, and from Equation (18) onwards, the analysis is confined to uncorrelated networks. The quantity~$\Theta(\lambda)$ in Equation~(3) of~\cite{pastor_PRL} is also based on the assumption of uncorrelatedness.}


The steady-state degree distribution of the scale-free graphs are obtained in~\cite{math_SF}. The expression is
\all{
n_k= \displaystyle \frac{2\beta(\beta+1)}{k(k+1)(k+2)}
.}{nk_steady_old}
The same function is obtained through the rate equation approach in~\cite{drog_nk}. We can use this expression to obtain the explicit form of~\eqref{pkl_uncorr_1} for the case of scale-free graphs. We will need~$\bar{k}$, which is found as follows
\al{
\bar{k}&= \sum_k k n_k = 2\beta(\beta+1) \sum_{k=\beta}^{\infty} \frac{1}{(k+1)(k+2)}
\nonumber \\
&= 2\beta(\beta+1) \sum_{k=\beta}^{\infty} \bigg[\frac{1}{k+1}-\frac{1}{k+2} \bigg] 
.}
This sum telescopes, and we find
\al{
\bar{k}=  2\beta(\beta+1) \left(\frac{1}{\beta+1} \right)= 2\beta 
.}
Substituting this into~\eqref{pkl_uncorr_1}, we find the uncorrelated approximation to the scale-free graphs for the conditional degree distribution
\all{
p_u(\ell |k)=\frac{\beta+1}{(\ell+1)(\ell+2)}
.}{pkl_uncorr}
One can easily check that  (the sum telescopes) this distribution  satisfies
\eq{
\sum_{\ell=\beta}^{\infty} p_u(\ell|k)=1
.}

The assumption of uncorrelatedness fails to emulate real-world networks which reportedly exhibit correlations and assortativity~\cite{assort,vazquez,maslov}.

\section{Linear Preferential Attachment: Conditional Degree Distribution} \label{sec1}
\subsection{Conditional  Degree Distribution}

Consider the network growth process at time~$t$. Time increments by one after each node addition. Each newly-introduced node connects to~$\beta$ existing nodes. Let~$N_k(t)$ denote the expected value of the number of nodes with degree~$k$ at time~$t$. Each new link that is formed by  the newly-added node attaches to  a node of degree~$k$ with the following probability: 
\all{
\displaystyle \frac{k N_k}{\sum_{k=1}^{\infty} k N_k }
.}{prob_pref_1}
The denominator is the sum of degrees of all nodes. It equals twice the number of links. Denote the number of links of the initial graph by~$L(0)$. Then the denominator of~\eqref{prob_pref_1} becomes~${2L(0)+2 \beta t}$.

Let~$N_{k \ell}(t)$ be the number of nodes with degree~$\ell$ who are connected to a parent of degree~$k$. When a new node is added, $N_{k \ell}$ can be incremented in of the following two ways. If  the parent node in a parent-child pair of~${(k-1,\ell)}$ receives a link, and if the child in a parent-child pair of~${(k,\ell-1)}$ receives a link. These events happen with probabilities~${\frac{\beta(k-1)N_{k-1,\ell}}{2L(0)+2 \beta t}}$ and~${\frac{\beta(\ell-1)N_{k,\ell-1}}{2L(0)+2 \beta t}}$, respectively. Also, $N_{k \ell}$ can be decremented if  either the parent or the child  in a parent-child pair of~${(k,\ell)}$ receive a link. These events happen with probabilities~${\frac{\beta k N_{k,\ell}}{2L(0)+2 \beta t}}$ for the parent and~${\frac{\beta \ell N_{k,\ell}}{2L(0)+2 \beta t}}$ for the child. The last case  to consider is that upon addition of each new node, if a node of degree~$k$ receives a link, then $N_{k \beta}$ increments by one. Combining these cases, we arrive at the rate equation which describes the time evolution of~$N_{k \ell}$. We have:
\all{
&N_{k{\ell}}(t+1)-N_{k{\ell}}(t)= \beta \Bigg[
\frac{(k-1)N_{k-1,{\ell}}(t)}{2L(0)+2   \beta t} 
\nonumber \\
&- \frac{k N_{k,{\ell}}(t)}{2L(0)+2   \beta t} + \frac{(\ell-1)N_{k,{\ell}-1}(t)}{2L(0)+2   \beta t}- \frac{\ell N_{k,{\ell}}(t)}{2L(0)+2   \beta t}
\nonumber \\ &
 + \frac{(k-1)N_{k-1}(t)}{2L(0)+2   \beta t} \delta(\ell,\beta)\Bigg]    .
}{N_diff_0}
By definition, we have:
\eq{
N_{k{\ell}}(t)= n_{k{\ell}}(t) [N(0)+t]
,}
so we have
\al{
&n_{k{\ell}}(t+1)[N(0)+t+1] -n_{k{\ell}}(t)[N(0)+t] \nonumber \\
&= \beta \Bigg[
\frac{(k-1)[N(0)+t] n_{k-1,{\ell}}(t)}{2L(0)+2   \beta t} - \frac{k[N(0)+t]  n_{k,{\ell}}(t)}{2L(0)+2   \beta t} 
\nonumber \\
&+ \frac{(\ell-1)[N(0)+t] n_{k,{\ell}-1}(t)}{2L(0)+2   \beta t}- \frac{\ell [N(0)+t] n_{k,{\ell}}(t)}{2L(0)+2   \beta t} \nonumber \\
 &+ \frac{(k-1)[N(0)+t] n_{k-1}(t)}{2L(0)+2   \beta t} \delta(\ell,\beta)\Bigg]   
.}
We focus on the steady-state limit of this equation. Let~${n_{k{\ell}}}$ denote ${n_{k{\ell}}(t)}$ in the limit of~${t \rightarrow \infty}$. Also note that in this limit, the following holds: 
\eq{
\lim_{t \rightarrow \infty} \displaystyle \frac{N(0)+t}{2L(0)+2   \beta t} = \frac{1}{2\beta}
}
So for the steady-state we have
\al{
&n_{k{\ell}} =  
\frac{(k-1)  n_{k-1,{\ell}} }{2} - \frac{k  n_{k,{\ell}}}{2} \nonumber \\
&+ \frac{(\ell-1) n_{k,{\ell}-1}}{2}- \frac{\ell   n_{k,{\ell}}}{2}
 + \frac{(k-1)  n_{k-1}}{2} \delta(\ell,\beta)    
.}
This can be rearranged and expressed as follows
\all{
(k+\ell+2)n_{k{\ell}}& =  
(k-1)  n_{k-1,{\ell}}  +  (\ell-1) n_{k,{\ell}-1}  \nonumber \\
&
+  (k-1)  n_{k-1}  \delta(\ell,\beta)    
.}{nkl_temp}
Using the expression for the  degree distribution in the steady-state as given in~\eqref{nk_steady_old}, we have
\eqq{
n_{k}= \frac{2\beta(\beta+1)}{k(k+1)(k+2)} u(k-\beta),
}{nk_inf}
where~$u(x)$ is the Heaviside step function; it is zero for negative inputs and is one otherwise. Using this, and also dividing all terms by the factor~${(k+\ell+2)}$, ~\eqref{nkl_temp} becomes

\all{
&n_{k{\ell}} =  
\frac{(k-1)}{k+\ell+2}  n_{k-1,{\ell}}   
+  \frac{(\ell-1)}{k+\ell+2} n_{k,{\ell}-1} \nonumber \\
&+  \frac{2\beta(\beta+1)}{k(k+1)(k+\beta+2)}  \delta(\ell,\beta)  u(k-1-\beta)   
.}{nkl_temp_2_text}

This difference equation is solved in Appendix~\ref{app:nkl_diff_solve}. The result is 

\all{
n_{k{\ell}}&= 2\beta^2(\beta+1)\frac{(k-1)!(\ell-1)!}{(k+\ell+2)!} 
\nonumber \\
& \times \sum_{\mu=\beta+1}^k \CC{\mu+\beta+1}{\beta}
\CC{k+\ell-\beta-\mu}{\ell-\beta}
.}{nkl_final_form}

To find the conditional degree distribution, note that we have
\all{
p(\ell   | k)= \frac{N_{k\ell}+N_{\ell k}}{k N_k}= 
\frac{n_{k\ell}+n_{\ell k}}{k n_k}
.}{p_from_n}
 Plugging the expression given in~\eqref{nk_inf} in the denominator and using~\eqref{nkl_final_form} for the terms in the numerator, for ${k,\ell \geq \beta}$ we get

\all{
p(\ell   | k)&= \beta (k+1)(k+2) \frac{(k-1)!(\ell-1)!}{(k+\ell+2)!} 
\nonumber \\
&\times \Bigg[ 
\sum_{\mu=\beta+1}^k \CC{\mu+\beta+1}{\beta}
\CC{k+\ell-\beta-\mu}{\ell-\beta}  \nonumber \\
&+
\sum_{\mu=\beta+1}^\ell \CC{\mu+\beta+1}{\beta}
\CC{k+\ell-\beta-\mu}{k-\beta} 
\Bigg] .
  }{pkl_final_1}

This can be equivalently expressed as follows:
\all{
p(\ell   | k)&= \frac{\beta}{k \ell \CC{k+\ell+2}{\ell}} 
\nonumber \\
& \times \Bigg[ 
\sum_{\mu=\beta+1}^k \CC{\mu+\beta+1}{\beta}
\CC{k+\ell-\beta-\mu}{\ell-\beta} 
\nonumber \\
&+ \sum_{\mu=\beta+1}^\ell \CC{\mu+\beta+1}{\beta}
\CC{k+\ell-\beta-\mu}{k-\beta} 
\Bigg] .
  }{pkl_final_1_2}
%
%
Now we find the closed form expression for these sums through speculation. In Table~\ref{terms_1},~${p(\ell |k)}$ is presented for the first few permitted values of~$k$ (recall that~$k \geq \beta$). We obtain the following closed form expression: 
\all{
p(\ell | k)= \displaystyle \frac{\beta(\beta+1) \CC{k+\ell+2}{\ell+1}-\beta(\beta+2) \CC{2\beta+2}{\beta} \CC{k+\ell-2\beta}{\ell-\beta}}{k \ell(\beta+1) \CC{k+\ell+2}{\ell}}.
}{plk_FIN_1}

%
%
%

 \begin{widetext}

\begin{table}[t]
\centering 
\begin{tabular}{| c |c |} 
\hline 
 $k $& $p(\ell | k ) $  \\ \hline
 $\beta+ 1$&  $\displaystyle \frac{ -\beta(\beta+2)(\ell-\beta+1) \CC{2\beta+2}{\beta} + \beta(\beta+1) \CC{\beta+\ell+3}{\ell+1}   }{(\beta+1)^2 \ell \CC{\beta+\ell+3}{\ell}}$ \\ \hline
$\beta+ 2$&  $\displaystyle  \frac{\frac{-\beta}{2} (\beta+2)(\ell-\beta+1)(\ell-\beta+2)\CC{2\beta+2}{\beta}+\beta(\beta+1)\CC{\beta+\ell+4}{\ell+1}}{(\beta+1)(\beta+2)\ell \CC{\ell+\beta+4}{\ell}} $ \\ \hline
$\beta+ 3$& $\displaystyle  \frac{\frac{-\beta}{6} (\beta+3)(\ell-\beta+1)(\ell-\beta+2) (\ell-\beta+3) \CC{2\beta+2}{\beta}+\beta(\beta+1)\CC{\beta+\ell+5}{\ell+1}}{(\beta+1)(\beta+3)\ell \CC{\ell+\beta+5}{\ell}}   $  
\\ \hline
$\beta+  4$&  $\displaystyle  \frac{\frac{-\beta}{24} (\beta+2)(\ell-\beta+1)(\ell-\beta+2) (\ell-\beta+3)(\ell-\beta+4) \CC{2\beta+2}{\beta}+\beta(\beta+1)\CC{\beta+\ell+6}{\ell+1}}{(\beta+1)(\beta+4)\ell \CC{\ell+\beta+6}{\ell}}   $  
  \\
 [1ex] 
\hline 
\end{tabular}
\caption{$p(\ell|k)$ for the first few values of~$k$. These expressions follow a clear pattern whose general closed form can be readily recognized and formulated in closed form for general~$(k,\ell)$ pair, as given in Equation~\eqref{plk_FIN_1}.} 
\label{terms_1} 
\end{table}

 \end{widetext}

 Rearranging the terms, the conditional degree distribution  further simplifies to the following expression: 
\all{
p(\ell | k)= \displaystyle \frac{\beta(k+2)}{k \ell (\ell+1)}
-\displaystyle \frac{\beta \CC{2\beta+2}{\beta+1}}{k \ell} 
\frac{\CC{k+\ell-2\beta}{\ell-\beta}}{\CC{k+\ell+2}{\ell}}
.}{plk_FIN_2}

In Appendix~\ref{app_unity} we confirm that these probabilities   add up to unity. In Appendix~\ref{app:sum_up} we prove that~\eqref{plk_FIN_2}  is equivalent to~\eqref{pkl_final_1_2}, through mathematical induction. 

\begin{figure}[t]
  \centering
  \includegraphics[width=1 \columnwidth]{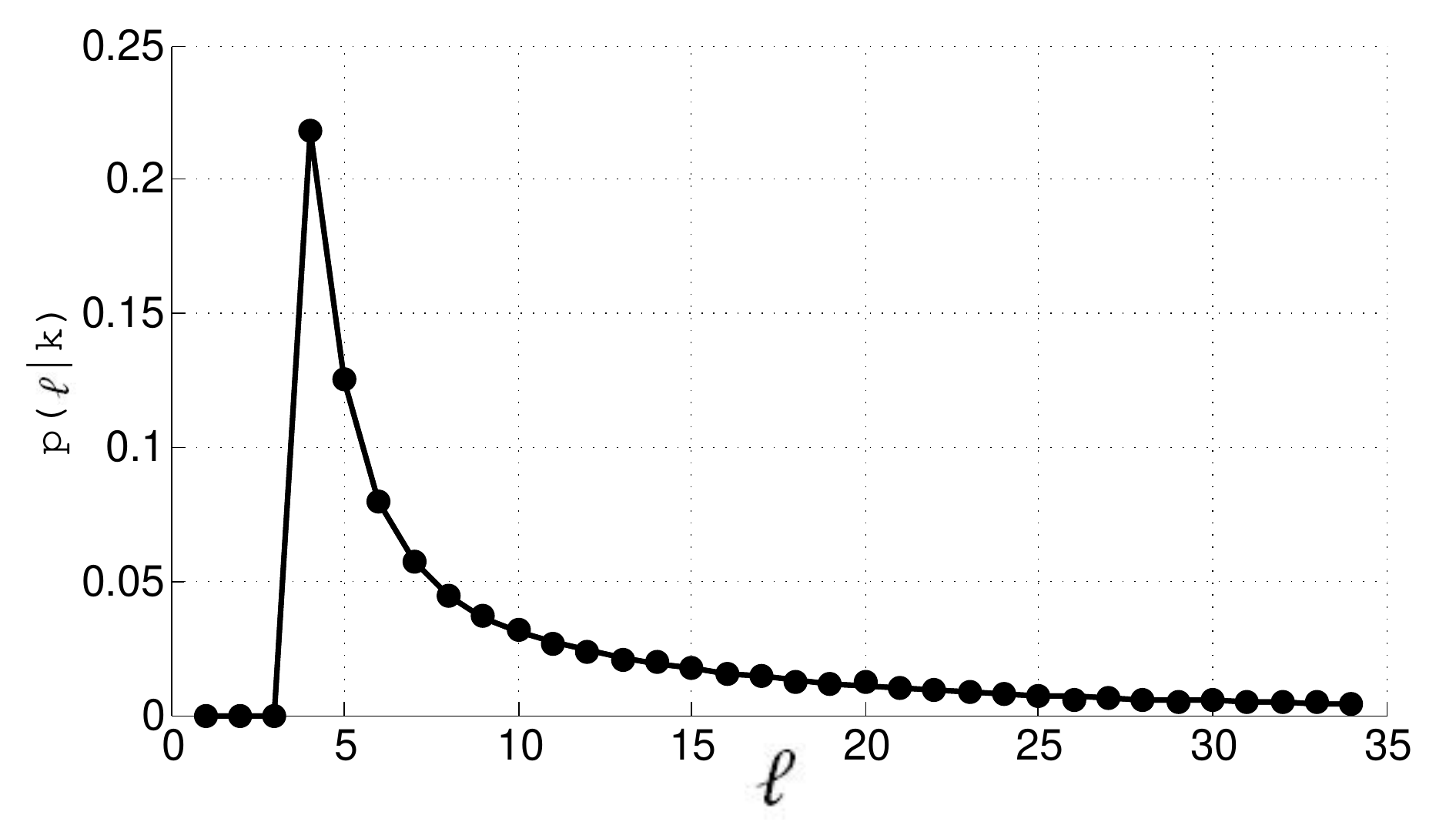}
  \caption[Figure ]%
  {The conditional degree distribution $p(\ell | k)$ as a function of $\ell$, as predicted by Equation ~\eqref{plk_FIN_2}, with the value of $k$ fixed at ${k=8}$. The growth process is preferential attachment with $\beta=4$. The number of initial nodes is 5,  and the distribution is obtained at $t=500$.  The results are averaged over 10 Monte Carlo simulations, and error bars are smaller than marker thickness.}
\label{fig_simul_1}
\end{figure}

 In Figure~\ref{fig_simul_1}, theoretical prediction is compared with simulations for a preferential attachment growth with ~$\beta=4$, with $N(0)=8$ after 500 timesteps.

\subsection{Joint Degree Distribution}
Now, using~\eqref{plk_FIN_2} together with~\eqref{pkl_def} we can also find the number of
links with incident edges of degrees~$k,\ell$. Multiplying~$p(\ell | k)$ by~$k N_k$ we have
\all{
L_{k,\ell}&=p(\ell | k) k N_k= 
 p(\ell | k)     \frac{2\beta(\beta+1)N}{ (k+1)(k+2)}
 \nonumber \\
 &=  
\displaystyle 
\frac{2 \beta^2(\beta+1) }{k(k+1) \ell (\ell+1)}N
\nonumber \\
&-\displaystyle \frac{2 \beta^2(\beta+1)}{k (k+1)(k+2)\ell} \CC{2\beta+2}{\beta+1}
\frac{\CC{k+\ell-2\beta}{\ell-\beta}}{\CC{k+\ell+2}{\ell}}N
.}{L_final_1}
This readily leads us to the joint degree distribution
\all{
p(k,\ell)&=\frac{L_{k,\ell}}{N \beta}=  
\displaystyle 
\frac{2 \beta(\beta+1) }{k(k+1) \ell (\ell+1)} \nonumber \\
&-\displaystyle \frac{2 \beta(\beta+1)}{k (k+1)(k+2)\ell} \CC{2\beta+2}{\beta+1}
\frac{\CC{k+\ell-2\beta}{\ell-\beta}}{\CC{k+\ell+2}{\ell}}
.}{L_final}
We can rearrange the factors to recast this equation into the following manifest-symmetric form: 
\all{
p(k,\ell)&= 
\displaystyle 
\frac{2 \beta(\beta+1) }{k(k+1) \ell (\ell+1)} \nonumber \\
&-\displaystyle \frac{2 \beta(\beta+1)}{k (k+1)\ell(\ell+1)} \CC{2\beta+2}{\beta+1}
\frac{\CC{k+\ell-2\beta}{\ell-\beta}}{\CC{k+\ell+2}{\ell+1}}
.}{joint_manifest}

We can relate this expression to the results obtained in~\cite{redner1} for the case of~${\beta=1}$. Using Equation~(41) in~\cite{redner1}, we have:
\al{
n_{k\ell}+n_{\ell k}&= 
\displaystyle \frac{  \frac{4(\ell-1)}{k(k+1)}+\frac{4(k-1)}{\ell(\ell+1)}}{(k+\ell)(k+\ell+1)(k+\ell+2)} \nonumber \\
&+ \displaystyle \frac{  \frac{12(\ell-1)}{k}+\frac{12(k-1)}{\ell}}{(k+\ell-1)(k+\ell)(k+\ell+1)(k+\ell+2)}
,}
which after taking the common denominator and simplifications transforms into
\all{
&n_{k\ell}+n_{\ell k}= 
4 \Bigg[k^4 + k^3(4\ell+2)-k^2+2k(4\ell^3-8\ell-2)
\nonumber \\
&+(\ell+2)(\ell+1)\ell(\ell-1) \left. \Bigg] \right/ \Bigg[ 
k(k+1)\ell(\ell+1) \nonumber \\
& \times (k+\ell-1)(k+\ell)(k+\ell+1)(k+\ell+2) \Bigg]
.}{redner_test}
Now we show that~\eqref{L_final} yields the same. For the special case 
of~$\beta=1$,~\eqref{plk_FIN_1} becomes
\all{
&p(\ell | k)k n_k = \displaystyle \frac{2 \frac{(k+\ell+2)!}{(\ell+1)!(k+1)!}-12 \frac{(k+\ell-2)! }{(\ell-1)!(k-1)!}}{2 k \ell  \frac{(k+\ell+2)!}{\ell ! (k+2)!}}
\frac{4k}{k(k+1)(k+2)}
\nonumber \\
&= 4 \displaystyle  \Bigg[   (k+\ell-1)(k+\ell)(k+\ell+1)(k+\ell+2)   
\nonumber \\
& - 6    k(k+1)  \ell(\ell+1) \left. \Bigg] \right/ \Bigg[   k (k+1) \ell (\ell+1)    \nonumber \\
& \times(k+\ell-1)(k+\ell)(k+\ell+1)(k+\ell+2)  \Bigg]
,}{my_test}
and through elementary algebraic steps one can show that
\al{
   &(k+\ell-1)(k+\ell)(k+\ell+1)(k+\ell+2)   \nonumber \\
& - 6    k(k+1)  \ell(\ell+1)  
   = k^4 + k^3(4\ell+2)
   \nonumber \\
&-k^2+2k(4\ell^3-8\ell-2)+(\ell+2)(\ell+1)\ell(\ell-1)
,}
which means that~\eqref{redner_test} and~\eqref{my_test} are identical. So our result agrees with that of~\cite{redner1} for the special case of~${\beta=1}$.

\section{Shifted Linear Preferential Attachment} \label{sec:shifted_sec}
  
Another class of network growth models are those involving \emph{redirection}.  In these models, incoming nodes randomly pick destinations amongst the existing nodes, but have the option of rather attaching to the parents of those nodes, with a constant probability. In~\cite{redner1} (see section~III-C, under Equation~(18) for discussion), it is shown that these models are equivalent to the shifted version of preferential attachment. In particular, if the redirection probability is denoted by~$r$, then these modes are equivalent to a preferential attachment mechanism, where the selection probability is~${k+(\frac{1}{r}-2)}$, which is a constant added to the degree.

In addition to redirection models, there is another family of network growth models that reduce to the shifted linear kernel family. If one is considering the growth problem over directed networks and the attachment probabilities are proportional to  a linear combination of out-degree and in-degree for each node, this can also be reformulated within the framework of shifted-linear preferential attachment (see~\cite{redner1}, section~III~C3, under Equation~(11) for discussion). If the probability of attaching to an existing node of in-degree~$k_i$, out-degree~$k_o$ and total degree~${k=k_i+k_o}$ is equal to~${a k_i + b k_o}$, then one can view this problem  as a shifted linear attachment model with probabilities equal to~$k+(\frac{b}{a}-\beta)$. 

In this section, we consider the problem of preferential attachment where the probability of attaching to an existing node of degree~$k$ is given by~${k+\theta}$, where~$\theta$ is a constant.

In this case, a node of degree~$k$ receives a link from a newly-born node with the following probability: 
\eq{
\frac{k+\theta}{\sum_{i:i\in V(t)} (k+\theta)}
.}
The denominator equals the denominator of the non-shifted case, which was~$2L(0)+\beta t$, plus~$\theta N(t) $. 

The rate equation for $N_{k \ell}(t)$ is obtained through identical steps undertaken  in Section~\ref{sec1}  which eventuated in~\eqref{N_diff_0}. In the case of shifted-linear kernel, the rate equation is
  
\al{
&N_{k{\ell}}(t+1)-N_{k{\ell}}(t)=  
\frac{\beta(k-1+\theta)N_{k-1,{\ell}}}{2L(0)+2   \beta t + \theta N(t)}
\nonumber \\
& - \frac{\beta(k+\theta) N_{k,{\ell}}}{2L(0)+2   \beta t+ \theta N(t)} + \frac{\beta(\ell-1+\theta)N_{k,{\ell}-1}}{2L(0)+2   \beta t+ \theta t}
\nonumber \\
&- \frac{\beta(\ell+\theta) N_{k,{\ell}}}{2L(0)+2   \beta t+ \theta N(t)} + \frac{\beta(k-1+\theta)N_{k-1}}{2L(0)+2   \beta t+ \theta N(t)} \delta(\ell,\beta)     
.}
Similar to the non-shifted version, we replace~$N_{k\ell}$ with~${[N(0)+t]n_{k \ell}}$. The left hand side becomes~$n_{k \ell}$ in the steady state. In this limit, we also have 
\eq{
\lim_{t \rightarrow \infty} \displaystyle \frac{N(0)+t}{2L(0)+2   \beta t+ \theta [N(0)+t]} = \frac{1}{2\beta+\theta}
.}
So for the steady-state we have
\al{
&n_{k{\ell}} =
\frac{(k-1+\theta)n_{k-1,{\ell}}}{2  + \frac{\theta}{\beta} } - \frac{(k+\theta) n_{k,{\ell}}}{2+\frac{\theta}{\beta} } 
\nonumber \\
&+ \frac{(\ell-1+\theta)n_{k,{\ell}-1}}{2+\frac{\theta}{\beta} }- \frac{(\ell+\theta) n_{k,{\ell}}}{2+\frac{\theta}{\beta} }
\nonumber \\
& + \frac{(k-1+\theta)n_{k-1}}{2+\frac{\theta}{\beta} } \delta(\ell,\beta)   
.}
This can be rearranged and expressed as follows
\all{
&\Big[k+\ell+2+\theta(2+\frac{1}{\beta})\Big] n_{k{\ell}} =  
(k-1+\theta)  n_{k-1,{\ell}}   \nonumber \\
&+  (\ell-1+\theta) n_{k,{\ell}-1} 
+  (k-1+\theta)  n_{k-1}  \delta(\ell,\beta)    
.}{nkl_temp_shifted}
To proceed, we need the degree distribution for the shifted linear preferential attachment. We solve for it in Appendix~\ref{app:nk_shifted}. Also let us define: 
\al{
\nu \stackrel{\text{def}}{=} 2+ \displaystyle \frac{\theta}{\beta}.
}
The degree distribution obtained in Appendix~\ref{app:nk_shifted}  is as follows:
\all{
n_k =  \displaystyle \frac{\nu}{\beta+\theta+\nu}  \displaystyle \frac{\Gamma(k+\theta)}{\Gamma(\beta+\theta)}
\displaystyle \frac{\Gamma (\beta+1+\theta+\nu ) }{\Gamma ( k+1+\theta+\nu )}
.}{deg_dist_shifted}
Substituting for~$n_{k-1}$ in~\eqref{nkl_temp_shifted} using this expression, and also dividing all terms by the factor~${(k+\ell+2\theta+\nu)}$, and also nothing that~${(k-1+\theta)\Gamma(k-1+\theta)=\Gamma(k+\theta)}$, then ~\eqref{nkl_temp_shifted} transforms into

\all{
& n_{k{\ell}} =  
\frac{(k-1+\theta)  n_{k-1,{\ell}} }{k+\ell+2\theta+\nu}  +  \frac{(\ell-1+\theta) n_{k,{\ell}-1} }{k+\ell+2\theta+\nu} \nonumber \\
&+ \left(\displaystyle \frac{\nu}{\beta+\theta+\nu}\right)   
\left( \displaystyle \frac{\Gamma (\beta+1+\theta+\nu )}{\Gamma(\beta+\theta)} \right)
\nonumber \\
& \times \frac{1 }{(k+\beta+2\theta+\nu)}  
\displaystyle \frac{\Gamma(k+\theta) }{\Gamma ( k+\theta+\nu )}
\delta(\ell,\beta)    
.}{nkl_temp_shifted_2_text}

This difference equation is solved in Appendix~\eqref{app:shifted_diff_solve}. The solution is

\all{
n_{k \ell} &= \displaystyle A \frac{\Gamma(k+\theta)  \Gamma(\ell+\theta)} 
{\Gamma(k+\ell+2\theta+\nu+1)} \nonumber \\
& \times 
\sum_{\mu=\beta+1}^k \displaystyle \frac{\Gamma(\mu+\beta+2\theta+\nu) }{\Gamma ( \mu+\theta+\nu )  } 
\CC{k+\ell-\beta-\mu}{\ell-\beta}
.}{shifted_nkl_use}

To constitute~$p(\ell|k)$ from this, we need to compute
\all{
p(\ell|k)= \displaystyle \frac{n_{k \ell}+n_{\ell k}}{k n_k }
.}{plk_ref_def}
Using the degree distribution given in~\eqref{deg_dist_shifted}, the denominator equals
\all{
k n_k &= \displaystyle  \nu   \displaystyle \frac{k \Gamma(k+\theta)}{\Gamma(\beta+\theta)}
\displaystyle \frac{\Gamma (\beta+\theta+\nu ) }{\Gamma ( k+1+\theta+\nu )}  
\nonumber \\
&=\displaystyle A \Gamma(\beta+\theta) 
\displaystyle \frac{ k \Gamma(k+\theta) }{\Gamma ( k+1+\theta+\nu )}
}{knk_shifted}
Using~\eqref{shifted_nkl_use}, the numerator of~\eqref{plk_ref_def} is
\al{
n_{k \ell}+n_{\ell k}  &= \displaystyle A \frac{\Gamma(k+\theta)  \Gamma(\ell+\theta)} 
{\Gamma(k+\ell+2\theta+\nu+1)} \nonumber \\
& \times \Bigg[ 
\sum_{\mu=\beta+1}^k \displaystyle \frac{\Gamma(\mu+\beta+2\theta+\nu) }{\Gamma ( \mu+\theta+\nu )  } 
\CC{k+\ell-\beta-\mu}{\ell-\beta}
\nonumber \\
&+ 
\sum_{\mu=\beta+1}^{\ell} \displaystyle \frac{\Gamma(\mu+\beta+2\theta+\nu) }{\Gamma ( \mu+\theta+\nu )  } 
\CC{k+\ell-\beta-\mu}{k-\beta}
\Bigg] 
.}
Hence the conditional degree distribution becomes
\all{
p(\ell|k)&=  
\displaystyle   \frac{\Gamma(k+1+\theta+\nu)  \Gamma(\ell+\theta)} 
{\Gamma(\beta+\theta) k \Gamma(k+\ell+2\theta+\nu+1)} \nonumber \\
& \times \Bigg[ 
\sum_{\mu=\beta+1}^k \displaystyle \frac{\Gamma(\mu+\beta+2\theta+\nu) }{\Gamma ( \mu+\theta+\nu )  } 
\CC{k+\ell-\beta-\mu}{\ell-\beta}
\nonumber \\
&+ 
\sum_{\mu=\beta+1}^{\ell} \displaystyle \frac{\Gamma(\mu+\beta+2\theta+\nu) }{\Gamma ( \mu+\theta+\nu )  } 
\CC{k+\ell-\beta-\mu}{k-\beta}
\Bigg] 
.}{shifted_FIN}

\begin{figure}[t]
  \centering
  \includegraphics[width=1 \columnwidth]{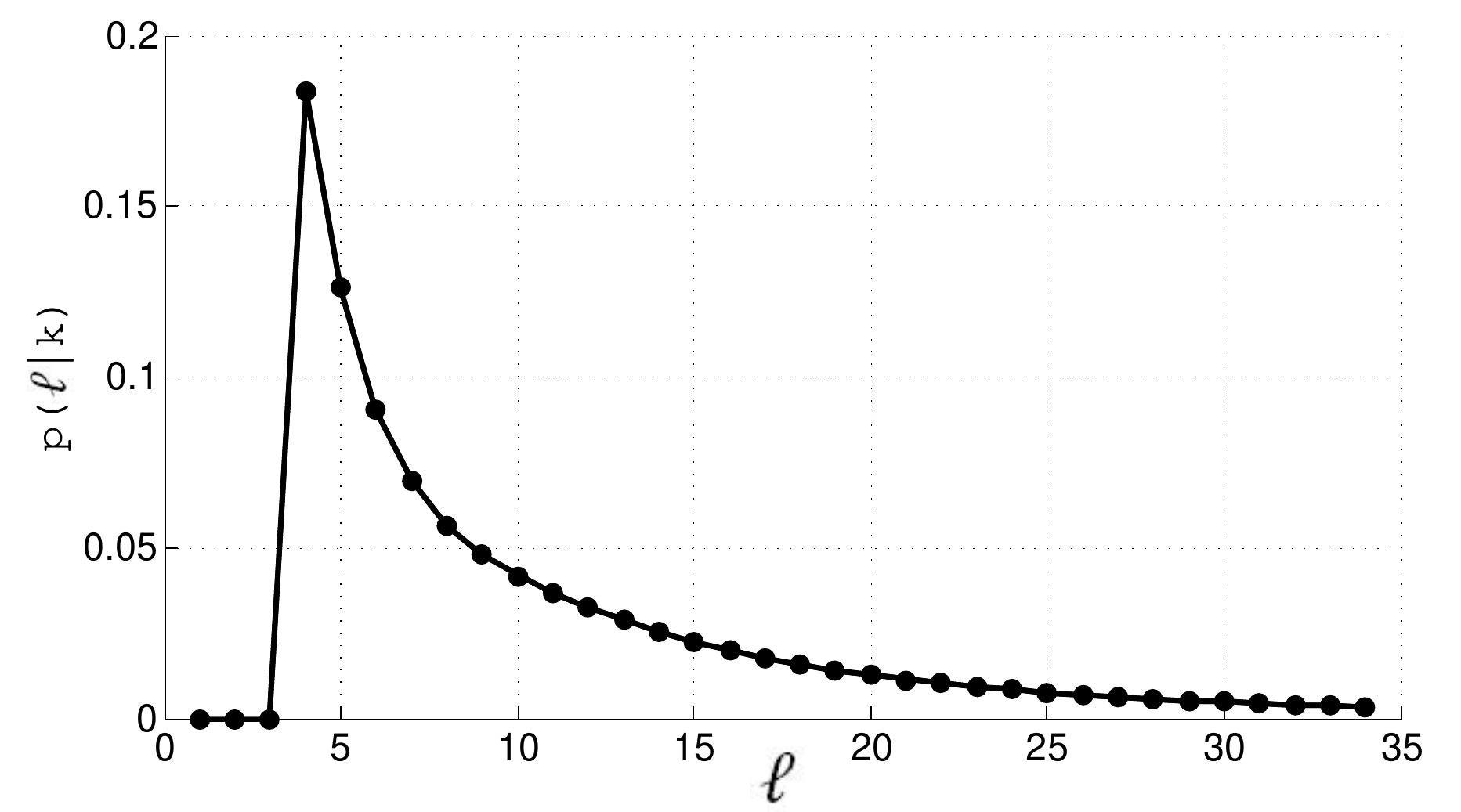}
  \caption[Figure ]%
  {The conditional degree distribution $p(\ell | k)$ as a function of $\ell$, as predicted by Equation ~\eqref{shifted_FIN}, with the value of $k$ fixed at ${k=8}$. The growth process is preferential attachment with $\beta=4$. The value of $\theta$ is 4.4. This means that a node with degree $x$ receives a link fro the incoming node with probability that is proportional to ${x+4.4}$. The number of initial nodes is 5,  and the distribution is obtained at $t=500$. }
\label{fig_simul_2}
\end{figure}

For the special case of~$\theta=0$, it is readily observable that one recovers~\eqref{pkl_final_1_2}.  
Analytical verification of ${\sum_{\ell} p(\ell|k)=1}$ is formidable in this case. But we numerically confirmed  this condition for different values of $\theta$, $\beta$ and $k$.

In the range~${\theta \ll \beta}$, which means slight deviation from the pure preferential attachment model, the expression in~\eqref{shifted_FIN} can be approximated as follows
\all{
p(\ell|k)=&  
\displaystyle   \frac{(\beta+\theta) \Gamma(k+1+\theta+\nu)   } 
{  k (\ell+\theta)(\ell+1+\theta)\Gamma(k+\ell+2\theta+\nu+1)} \nonumber \\
& \times \Bigg[ \displaystyle  \frac{\Gamma(k+\ell+3+\theta)}{\Gamma(k+2+\theta)}
\nonumber \\
&-\CC{k+\ell-2\beta}{\ell-\beta}\frac{\Gamma(\ell+2+\theta) \Gamma(3+2\beta+2\theta)}{[\Gamma(\beta+2+\theta)]^2} \Bigg] 
.}{pkl_asymp_shifted}

Similar to the non-shifted case, we can extract the joint degree distribution using the relation
\eq{
p(k,\ell)= \frac{k n_k}{\beta}  p(\ell|k).
}

Using  the expressions obtained in~\eqref{knk_shifted} and~\eqref{shifted_FIN}, we arrive at
\all{
p(k,\ell)&=  
\displaystyle   \frac{\nu \Gamma(k+\theta)  \Gamma(\ell+\theta) \Gamma(\beta+\theta+\nu)} 
{\beta \big[\Gamma(\beta+\theta) \big]^2 \Gamma(k+\ell+2\theta+\nu+1)} \nonumber \\
& \times \Bigg[ 
\sum_{\mu=\beta+1}^k \displaystyle \frac{\Gamma(\mu+\beta+2\theta+\nu) }{\Gamma ( \mu+\theta+\nu )  } 
\CC{k+\ell-\beta-\mu}{\ell-\beta}
\nonumber \\
&+ 
\sum_{\mu=\beta+1}^{\ell} \displaystyle \frac{\Gamma(\mu+\beta+2\theta+\nu) }{\Gamma ( \mu+\theta+\nu )  } 
\CC{k+\ell-\beta-\mu}{k-\beta}
\Bigg] 
.}{p_joint_shifted}

Note that this expression is manifest-symmetric, as one would expect for the joint degree distribution.

\section{Summary and Discussion}

Preferential attachment models of network growth are of particular interest of network science scholars because they give rise to power-law degree distributions which is observed empirically in various contexts. In this paper, we have considered the general case where each new node connects to~$\beta$ existing nodes. We have employed the rate equation approach to describe the evolution of~$n_{k \ell}$, which is the fraction of nodes of degree~$\ell$ who are connected to a parent node of degree~$k$. After solving the rate equation, we have extracted the  conditional degree distribution, denoted by~$p(k|\ell)$, which is the degree distribution of the neighbors of a node of degree~$k$. We have also obtained expressions for the joint degree distribution~$p(k,\ell)$. 

%

We have also considered the problem of shifted-linear preferential attachment, and described how a variety of classes of network growth models can be reformulated under this overarching framework. We have found expressions for the conditional and joint degree distributions in the form of functions of a finite sum, unlike the non-shifted case where the solution was obtained in closed form. 

\appendix

\section{Solving the Difference Equation~\eqref{nkl_temp_2_text}} \label{app:nkl_diff_solve}

Let us repeat the difference equation we want to solve for easy reference:
\all{
&n_{k{\ell}} =  
\frac{(k-1)}{k+\ell+2}  n_{k-1,{\ell}}   
+  \frac{(\ell-1)}{k+\ell+2} n_{k,{\ell}-1} \nonumber \\
&+  \frac{2\beta(\beta+1)}{k(k+1)(k+\beta+2)}  \delta(\ell,\beta)  u(k-1-\beta)   
.}{nkl_temp_2}

Now define
\all{
m_{k{\ell}} \stackrel{\text{def}}{=} \frac{(k+\ell+2)!}{(k-1)! (\ell-1)!} n_{k{\ell}}
.}{mkl_def}
The following relations hold
\al{
\begin{cases}
\displaystyle  n_{k,{\ell}}
= \displaystyle  \frac{(k-1)! (\ell-1)!}{(k+\ell+2)!} m_{k,{\ell}} \\ \\
\displaystyle (k-1) n_{k-1,{\ell}}
= \displaystyle  \frac{(k-1)! (\ell-1)!}{(k+l+1)!} m_{k-1,{\ell}} \\ \\
\displaystyle (\ell-1) n_{k,{\ell}-1}
= \displaystyle  \frac{(k-1)! (\ell-1)!}{(k+l+1)!} m_{k,{\ell}-1}.
\end{cases}
}
Plugging these into~\eqref{nkl_temp_2} and multiplying through by 
the factor~${\frac{(k+\ell+2)!}{(k-1)! (\ell-1)!}}$ we obtain
\all{
m_{k{\ell}} &=  
m_{k-1,{\ell}}   
+ m_{k,{\ell}-1} \nonumber \\
&+  \frac{2\beta(\beta+1)}{k(k+1)(k+\beta+2)} 
\frac{(k+\beta+2)!}{(k-1)! (\beta-1)!}
\nonumber \\
&\times 
 \delta(\ell,\beta)  u(k-1-\beta)   
.}{nkl_temp_3}
Note that the following holds
\al{
&\frac{2\beta(\beta+1)}{k(k+1)(k+\beta+2)} 
\frac{(k+\beta+2)!}{(k-1)! (\beta-1)!} \nonumber \\
&  =  2\beta(\beta+1)  
\frac{(k+\beta+1)!}{(k+1)! (\beta-1)!}
 = 
  2\beta^2(\beta+1)  \CC{k+\beta+1}{\beta}
.}
So we arrive at the following difference equation
\all{
m_{k{\ell}}& =  
m_{k-1,{\ell}}   
+ m_{k,{\ell}-1} \nonumber \\
&+    2\beta^2(\beta+1)  \CC{k+\beta+1}{\beta}
 \delta(\ell,\beta)  u(k-1-\beta)   
.}{mkl_diff}
To solve this partial difference equation, we use the two-dimensional Z-transform defined as follows:
\al{
\psi(z,y) \stackrel{\text{def}}{=} \displaystyle 
\sum_{k=0}^{\infty} \sum_{\ell=0}^{\infty} m_{k,\ell} z^{-k} y^{-\ell}
.}
Multiplying~\eqref{mkl_diff} through by~${z^{-k} y^{-\ell}}$ and summing over~${k,\ell}$ we obtain
\al{
\psi(z,y)&=z^{-1} \psi(z,y)+y^{-1} \psi(z,y) \nonumber \\
&+  2\beta^2(\beta+1) \sum_{\mu=\beta+1}^{\infty} \CC{\mu+\beta+1}{\beta} z^{-\mu} y^{-\beta}
.}
This can be rearranged to yield
\all{
\psi(z,y)=\frac{2\beta^2(\beta+1)}{1-z^{-1}-y^{-1}} \sum_{\mu=\beta+1}^{\infty} \CC{\mu+\beta+1}{\beta} z^{-\mu} y^{-\beta}.
 }{psi_1}
To take the inverse Z-transform, we must evaluate the following integral:
\al{
m_{k\ell} = \frac{-1}{4\pi^2} \displaystyle \oint_z \oint_y \psi(z,y) z^{k-1} y^{\ell-1} dz dy
.}
Using  the expression for~${\psi(z,y)}$ from~\eqref{psi_1}, we get
\all{
m_{k\ell} = 2\beta^2(\beta+1) \frac{-1}{4\pi^2} \displaystyle 
\sum_{\mu=\beta+1}^{\infty} \CC{\mu+\beta+1}{\beta} 
I_{k \ell}
,}{m_I}
where we have defined
\al{
I_{k \ell} \stackrel{\text{def}}{=} 
\frac{-1}{4\pi^2} \oint_z \oint_y 
\frac{z^{-{\mu}} y^{-\beta}  z^{k-1} y^{\ell-1} }{1-z^{-1}-y^{-1}}
  dz dy
.}  
This can be equivalently expressed as follows:
 \al{
I_{k \ell} = 
\frac{-1}{4\pi^2} \oint_z \oint_y 
\displaystyle \frac{z^{k-{\mu}} y^{\ell-\beta} }{z-\left( \frac{y}{y-1}\right) } \frac{1}{y-1}  dz dy
.}
First we integrate over~$z$. There is an ordinary pole at~${z=\frac{y}{y-1}}$, so we get   
\al{
I_{k \ell} &=
\frac{1}{2\pi i} \oint_y 
\displaystyle \left(\frac{y}{y-1}\right)^{k-{\mu}} \frac{y^{\ell-\beta}}{y-1}   dy
\nonumber \\
&=\frac{1}{2\pi i} \oint_y 
\displaystyle \frac{y^{k-\mu+\ell-\beta}}{(y-1)^{k-\mu+1}}   dy
.}
The pole at~$y=1$ is of order~${k-\mu+1}$, so we have 
  \all{
I_{k \ell}
&=\frac{1}{(k-\mu)!} \frac{d^{k-\mu}}{dy^{k-\mu}} \left. y^{k-\mu+\ell-\beta}
\right|_{y=1}
\nonumber \\
&= \frac{(k-\mu+\ell-\beta)!}{(k-\mu)!(\ell-\beta)!}= \CC{k-\mu+\ell-\beta}{\ell-\beta}
.}{I_result}
Plugging this result in~\eqref{m_I} we obtain
\al{
m_{k{\ell}}= 2\beta^2(\beta+1) \sum_{\mu=\beta+1}^k \CC{\mu+\beta+1}{\beta}
\CC{k-\mu+\ell-\beta}{\ell-\beta}.
}
Now, using~\eqref{mkl_def}, we arrive at
\all{
n_{k{\ell}}&= 2\beta^2(\beta+1)\frac{(k-1)!(\ell-1)!}{(k+\ell+2)!} 
\nonumber \\
& \times \sum_{\mu=\beta+1}^k \CC{\mu+\beta+1}{\beta}
\CC{k+\ell-\beta-\mu}{\ell-\beta}
.}{nkl_final_form_app}


\section{Proof of Conditional Degree Distribution~\eqref{plk_FIN_2}  Adding Up to Unity} \label{app_unity}
We want to prove the following holds
\all{
\displaystyle \sum_{\ell=\beta} ^{\infty} \left[  \displaystyle \frac{\beta(k+2)}{k \ell (\ell+1)}
-\displaystyle \frac{\beta \CC{2\beta+2}{\beta+1}}{k \ell} 
\frac{\CC{k+\ell-2\beta}{\ell-\beta}}{\CC{k+\ell+2}{\ell}} \right] =1
.}{pkl_temp_1}
The first sum is trivial. Using the fact that
\al{
\displaystyle \frac{1}{ \ell (\ell+1)}= \displaystyle \frac{1}{ \ell }-\displaystyle \frac{1}{ \ell+1}
,}
then ${\displaystyle \sum   \displaystyle \frac{\beta(k+2)}{k \ell (\ell+1)}}$ telescopes, and only the first term remains uncanceled. So we get
\all{
\displaystyle \sum_{\ell=\beta} ^{\infty}    \displaystyle \frac{\beta(k+2)}{k \ell (\ell+1)}=
   \displaystyle \frac{ k+2}{k }= 1+  \displaystyle \frac{ 2}{k }
}{sum_1}
So to prove that~\eqref{pkl_temp_1} holds, it suffices to show that
\all{
\displaystyle \sum_{\ell=\beta} ^{\infty} 
\displaystyle \frac{\beta \CC{2\beta+2}{\beta+1}}{k \ell} 
\frac{\CC{k+\ell-2\beta}{\ell-\beta}}{\CC{k+\ell+2}{\ell}}  = \displaystyle \frac{ 2}{k }
.}{sum_2}
It is equivalent to the following
\all{
\displaystyle \sum_{\ell=\beta} ^{\infty} 
\displaystyle \frac{1}{  \ell} 
\frac{\CC{k+\ell-2\beta}{\ell-\beta}}{\CC{k+\ell+2}{\ell}}  = \displaystyle \frac{ 2}{\beta }
\displaystyle  \frac{1}{ \CC{2\beta+2}{\beta+1} }
.}{sum_3}
We begin by using the following identity for the reciprocal of a binomial coefficient (see~\cite{binom_paper_1,binom_paper_2}) 
\all{
\displaystyle \frac{1}{\CC{k+\ell+2}{\ell}}=
(k+\ell+3) \displaystyle \int_0^1 t^{\ell} (1-t)^{k+2} dt.
}{binom_inv}
If we plug this expression in the left hand side of~\eqref{sum_3}, and exchange the order of summation and integration operations, we arrive at the following sum for the left hand side of~\eqref{sum_3}:
\all{
  \displaystyle \int_0^1  (1-t)^{k+2} \displaystyle \sum_{\ell=\beta} ^{\infty} 
\displaystyle \frac{ k+\ell+3}{  \ell}
 \CC{k+\ell-2\beta}{\ell-\beta} t^\ell  dt
.}{sum_4}

Hereafter we will denote to this term by $\Psi$, and the objective is to prove that 
\all{
\Psi= \displaystyle  \frac{2}{\beta}  \displaystyle  \frac{1}{ \CC{2\beta+2}{\beta+1} }
.}{PSI_obj}
Now let us focus on the summation that appears inside the integral. We intend to perform this summation first, whose result will be a function of~$t$, and then computing the integral on $t$ from zero to one. Let us define
\al{
\begin{cases}
\sigma \stackrel{\text{def}}{=}  \displaystyle \sum_{\ell=\beta} ^{\infty} 
\displaystyle \frac{ k+\ell+3}{  \ell}
 \CC{k+\ell-2\beta}{\ell-\beta} t^\ell  \\ \\
\sigma_1 \stackrel{\text{def}}{=}  \displaystyle \sum_{\ell=\beta} ^{\infty} 
\displaystyle  \CC{k+\ell-2\beta}{\ell-\beta} t^\ell  \\ \\
\sigma_2 \stackrel{\text{def}}{=}  \displaystyle \sum_{\ell=\beta} ^{\infty} 
 \CC{k+\ell-2\beta}{\ell-\beta}  \displaystyle \frac{t^\ell }{  \ell}.
\end{cases}
}
Note that the following holds
\al{ 
\sigma=\sigma_1+(k+3) \sigma_2
.}

So~$\Psi$  can be expressed in the following form: 
\all{
\Psi=   \displaystyle \int_0^1  (1-t)^{k+2} \bigg[ \sigma_1 + (k+3) \sigma_2 \bigg] dt
.}{PSI_1}
First we focus on~$\sigma_1$. Defining  the binomial coefficient to be zero when the lower index is greater than the upper index, we can drop the summation bounds and assume that all summations are from~${-\infty}$ to~${+\infty}$. 
 We can change the lower index of the binomial coefficient and write~$\sigma_1$ as follows
\al{
\sigma_1 =  \displaystyle \sum_{\ell}  
\displaystyle  \CC{k+\ell-2\beta}{k-\beta} t^\ell .
}
This can be  expressed equivalently in the following form
\al{
\sigma_1 =  t^{2 \beta-k} \displaystyle \sum_{\mu} 
\displaystyle  \CC{\mu}{k-\beta} t^\mu .
}
We use the following elementary identity which can be readily proved by Taylor series arguments:
\all{
\sum_{\mu} 
\displaystyle  \CC{\mu}{\nu} t^\mu = 
\displaystyle  \frac{t^{\nu}}{(1-t)^{\nu+1}}
.}{taylor_1}
So we obtain
\al{
\sigma_1 =  t^{2 \beta-k} \displaystyle  \frac{t^{k-\beta}}{(1-t)^{k-\beta+1}}
,}
which can be simplified to
\all{
\sigma_1 =  \displaystyle  \frac{t^{\beta}}{(1-t)^{k-\beta+1}}
.}{sigma_1_final}
Now let us focus on~$\sigma_2$. Changing the lower index of the binomial coefficient, it takes the following form
\al{
\sigma_2   = \displaystyle \sum_{\ell=\beta} ^{\infty} 
 \CC{k+\ell-2\beta}{k-\beta}  \displaystyle \frac{t^\ell }{  \ell}.
}
We remove the $\ell$ factor in the denominator by writing this sum in the following  form
\al{
\sigma_2 = \displaystyle \int^t  \displaystyle \sum_{\ell=\beta} ^{\infty} 
 \CC{k+\ell-2\beta}{k-\beta}  \displaystyle t^{\ell-1}  dt
.}
This can be recast as
\al{
\sigma_2 = \displaystyle \int^t  t^{2\beta-k-1} \displaystyle \sum_{\mu} 
 \CC{\mu}{k-\beta}  \displaystyle t^{\mu}  dt 
.}
Using the identity~\ref{taylor_1}, this becomes
\al{
\sigma_2  &= 
\displaystyle \int^t  t^{2\beta-k-1}  \displaystyle  \frac{t^{k-\beta}}{(1-t)^{k-\beta+1}} dt 
\nonumber \\
&= 
\displaystyle \int^t   \displaystyle  \frac{t^{\beta-1}}{(1-t)^{k-\beta+1}} dt \nonumber \\
.}
This indefinite integral is intricate to compute directly. We find a way to circumvent it. 
Looking back at~\eqref{sigma_1_final}, note that we can recast this as follows:
\al{
\sigma_2= \displaystyle \int^t \displaystyle \frac{\sigma}{t} dt
.}
Taking the time derivative from both sides, we get
\all{
\dot{\sigma_2} =  \displaystyle \frac{\sigma}{t}
.}{sigma_dot}
Now let us get back to~\eqref{PSI_1}. Integrating by parts,   we have
\all{
\Psi &= \left. -\displaystyle \frac{(1-t)^{k+3}}{k+3}  \bigg[ \sigma_1 + (k+3) \sigma_2 \bigg]  \right|_{t=0}^{t=1}
 \nonumber \\
&+ \displaystyle \frac{1}{k+3}  \displaystyle \int_0^1  (1-t)^{k+3} \bigg[ \dot{\sigma}_1 + (k+3)\dot{ \sigma}_2 \bigg] dt
}{PSI_2}
Note that the first term vanishes at both bounds. At $t=1$ the $(1-t)^{k+3}$ factor vanishes and 
at $t=0$ both $\sigma_1$ and $\sigma_2$ vanish. Also, from~\eqref{sigma_dot} we can replace the derivative of $\sigma_2$ and express it in terms of $\sigma_1$. The result is
\all{
\Psi =  \displaystyle \frac{1}{k+3}  \displaystyle \int_0^1  (1-t)^{k+3} \bigg[ \dot{\sigma}_1 + (k+3)
  \displaystyle \frac{\sigma_1}{t} \bigg] dt.
}{PSI_3}
The expression for $\sigma_1$ is given in~\eqref{sigma_1_final}. After differentiation and elementary algebraic steps, one can readily verify that the following holds: 
\al{
 \dot{\sigma}_1 + (k+3)  \displaystyle \frac{\sigma_1}{t}= 
 \displaystyle \frac{(k+\beta+3) t^{\beta-1} -2(\beta+1)t^{\beta}      }{(1-t)^{k+2-\beta}} 
 .
}
Plugging this into~\eqref{PSI_3}, we obtain
\all{
\Psi =  \displaystyle \frac{1}{k+3}  \displaystyle \int_0^1  (1-t)^{\beta+1}
\bigg[ (k+\beta+3) t^{\beta-1} -2(\beta+1)t^{\beta} \bigg] 
}{PSI_4}
Let us now define
\al{
\begin{cases}
\xi_1 \stackrel{\text{def}}{=}  \displaystyle \int_0^1  t^{\beta-1} (1-t)^{\beta+1}   \\ \\
\xi_2 \stackrel{\text{def}}{=}  \displaystyle \int_0^1   t^{\beta}   (1-t)^{\beta+1}  .
\end{cases}
}
So we can express $\Psi$ in the following compact form
\all{
\Psi= \displaystyle \frac{(k+\beta+3) \xi_1 - 2 (\beta+1) \xi_2 }{k+3}
.}{PSI_5}
The integrals which constitute $\xi_1, \xi_2$ are Euler\rq{}s integrals of first kind. Generally, for integers $x,y$, one defines the Beta function as: 
\al{
B(x,y)= \displaystyle \int_0^1 t^{x-1} (1-t)^{y-1} dt
.}
Integrating by parts, one can readily see that the Beta function satisfies
\al{
B(x+1,y)=  \displaystyle \frac{x}{x+y} B(x,y).
}
Using the basic properties of factorials, we immediately recognize
\al{
B(x,y)=   \displaystyle \int_0^1 t^{x-1} (1-t)^{y-1} dt = \displaystyle \frac{(x-1)! (y-1)!}{(x+y-1)!} .
}
Using this result, we obtain
\al{
\begin{cases}
\xi_1 =    \displaystyle \frac{(\beta-1)! (\beta+1)!}{(2\beta+1)!}               \\ \\
\xi_2=  \displaystyle \frac{(\beta)! (\beta+1)!}{(2\beta+2)!}  .
\end{cases}
}
One can easily verify that $\xi_1$ and $\xi_2$ satisfy 
\al{
(k+\beta+3) \xi_1 - 2 (\beta+1) \xi_2  = \displaystyle \frac{(k+3) (\beta-1)! (\beta+1)!}{(2\beta+1)!}
.}
Substituting this expression for the numerator of $\Psi$ in~\eqref{PSI_5} we obtain
\all{
\Psi=  \displaystyle \frac{ (\beta-1)! (\beta+1)!}{(2\beta+1)!}
.}{PSI_6}
Now, note that the following is true: 
\al{
\begin{cases}
(\beta-1)!=   \displaystyle \frac{(\beta+1)!}{\beta (\beta+1)} \\ \\
\displaystyle \frac{1}{(2\beta+1)!}=   \displaystyle \frac{2(\beta+1)}{(2\beta+2)!} \\ \\
\end{cases}
}
Plugging these two into~\eqref{PSI_6}, we arrive at
\al{
\Psi= \frac{2}{\beta} \displaystyle \frac{(\beta+1)! (\beta+1)!}{(2\beta+2)!} =  \displaystyle  \frac{2}{\beta}  \displaystyle  \frac{1}{ \CC{2\beta+2}{\beta+1} }
.}
Hence we have proven~\eqref{PSI_obj}, which means that the probabilities given in~\eqref{pkl_temp_1} do add up to unity.












\section{Proof of Equivalence of~\eqref{plk_FIN_2} and~\eqref{pkl_final_1_2}} \label{app:sum_up}

The objective is to show that
\all{
&\displaystyle \frac{\beta(k+2)}{k \ell (\ell+1)}
-\displaystyle \frac{\beta \CC{2\beta+2}{\beta+1}}{k \ell} 
\frac{\CC{k+\ell-2\beta}{\ell-\beta}}{\CC{k+\ell+2}{\ell}} \nonumber \\
&= \frac{\beta}{k \ell \CC{k+\ell+2}{\ell}} 
 \Bigg[ 
\sum_{\mu=\beta+1}^k \CC{\mu+\beta+1}{\beta}
\CC{k+\ell-\beta-\mu}{\ell-\beta} 
\nonumber \\
&+ \sum_{\mu=\beta+1}^\ell \CC{\mu+\beta+1}{\beta}
\CC{k+\ell-\beta-\mu}{k-\beta} 
\Bigg] 
.}{objective_sum_0}
Multiplying both sides by the factor~${k \ell \CC{k+\ell+2}{\ell}}$ and dividing all terms by $\beta$,  we get the following equivalent form for this equation

\all{
&\displaystyle \frac{ (k+2) \CC{k+\ell+2}{\ell} }{ (\ell+1)}
-\displaystyle  \CC{2\beta+2}{\beta+1} 
 \CC{k+\ell-2\beta}{\ell-\beta}  \nonumber \\
&= 
\sum_{\mu=\beta+1}^k \CC{\mu+\beta+1}{\beta}
\CC{k+\ell-\beta-\mu}{\ell-\beta} 
\nonumber \\
&+ \sum_{\mu=\beta+1}^\ell \CC{\mu+\beta+1}{\beta}
\CC{k+\ell-\beta-\mu}{k-\beta} 
.}{objective_sum_1}

We will prove that the two sides of~\eqref{objective_sum_1} are equivalent using mathematical induction for $\beta$. 
The initial step is to prove that the equality holds for the case of $\beta=1$. In this case, we have the following equality to prove

\all{
&\displaystyle \frac{ (k+2) \CC{k+\ell+2}{\ell} }{ (\ell+1)}
-\displaystyle  \CC{4}{2} 
 \CC{k+\ell-2}{\ell-1}  \nonumber \\
&= 
\sum_{\mu=2}^k \CC{\mu+2}{1}
\CC{k+\ell-1-\mu}{\ell-1} 
\nonumber \\
&+ \sum_{\mu=2}^\ell \CC{\mu+2}{1}
\CC{k+\ell-1-\mu}{k-1} 
.}{initial_1}

Let us first focus on the first sum on the right hand side. We define
\eqq{
S_1 \stackrel{\text{def}}{=} \displaystyle \sum_{\mu=2}^k \CC{\mu+2}{1}
\CC{k+\ell-1-\mu}{\ell-1} 
.}{s1_def}

Denoting $\mu+2$ by $\nu$, we have
\all{
S_1= 
 \displaystyle \sum_{\nu=4}^{k+2} \CC{\nu}{1}
\CC{k+\ell+1-\nu}{\ell-1} 
.}{s1_temp_1}
Similar to Appendix~\ref{app_unity}, we can assume that the upper summation   bounds are infinity, since the binomial coefficient vanishes for ${\nu>k+2}$. Using this,~\eqref{s1_temp_1}  can be equivalently written as follows
\all{
S_1= 
 \displaystyle \sum_{\nu=0}^{\infty} \CC{\nu}{1}
\CC{k+\ell+1-\nu}{\ell-1}  -  \displaystyle \sum_{\nu=0}^{3} \CC{\nu}{1}
\CC{k+\ell+1-\nu}{\ell-1}
.}{s1_temp_2}
The second sum is finite and has only three nonzero terms. For the first  sum, note that it has the form of a convolution on the upper index of the binomial coefficients, and remember that when two power series are multiplied together, the coefficient of the product is the convolution of their individual coefficients. In this case, we have  the coefficient of  ${x^{k+\ell+1}}$  in the multiplication of two series. The  coefficient of $x^{\nu}$ for one of the series is $\CC{\nu}{1}$, and  the  coefficient of $x^{\nu}$  for the other one is $\CC{\nu}{\ell-1}$. So we have the following power series multiplication for the first sum on the right hand side :
\all{
 \displaystyle \sum  \CC{\nu}{1} x^{\nu} 
\times  \displaystyle \sum  \CC{\nu}{\ell-1} x^{\nu}  
}{s1_temp_3}
The limits of the summations are implied by the binomial coefficients, as discussed above.  
From~\eqref{taylor_1}, we know that
\all{
\begin{cases}
\displaystyle \sum  \CC{\nu}{1} x^{\nu} =  \displaystyle \frac{x}{(1-x)^2}
\\ \\
 \displaystyle \sum  \CC{\nu}{\ell-1} x^{\nu} =  \displaystyle \frac{x^{\ell-1}}{(1-x)^{\ell}}
\end{cases}
}{s1_temp_4}
So we need to find the coefficient of~${x^{k+\ell+1}}$ in the following power series
\al{
\displaystyle \frac{x^{\ell}}{(1-x)^{\ell+2}}
.}
This is identical to the coefficient of~${x^{k+\ell+2}}$  in the following series:
\al{
\displaystyle \frac{x^{\ell+1}}{(1-x)^{\ell+2}}
,}
which is $\CC{k+\ell+2}{\ell+1}$, using~\eqref{taylor_1}. Plugging this result in~\eqref{s1_temp_2}, we get
\all{
S_1= 
 \displaystyle \CC{k+\ell+2}{\ell+1} -  \displaystyle \sum_{\nu=0}^{3} \CC{\nu}{1}
\CC{k+\ell+1-\nu}{\ell-1}
.}{s1_temp_5}
Now we insert this in~\eqref{initial_1}. Noting that the second sum on the right hand side of~\eqref{initial_1} is the same as the first sum but with $\ell$ and $k$ swapped, we arrive at the following equivalent form for~\eqref{initial_1}, which we have to prove as  the initial step of induction

\all{
&\displaystyle \frac{ (k+2) \CC{k+\ell+2}{\ell} }{ (\ell+1)}
-\displaystyle  \CC{4}{2} 
 \CC{k+\ell-2}{\ell-1}  = 
\displaystyle \CC{k+\ell+2}{\ell+1}+  
\displaystyle \CC{k+\ell+2}{k+1}
\nonumber \\
& -  
\displaystyle \sum_{\nu=0}^{3} \CC{\nu}{1}
\CC{k+\ell+1-\nu}{\ell-1}
- 
\displaystyle \sum_{\nu=0}^{3} \CC{\nu}{1}
\CC{k+\ell+1-\nu}{k-1}
.}{initial_3}
Now note that the first two terms on the right hand side are identical. Also it can be easily checked that the following is true:
\al{
\displaystyle \frac{ (k+2) \CC{k+\ell+2}{\ell} }{ (\ell+1)}= 
\displaystyle \CC{k+\ell+2}{k+1}.
}
Using this and noting that ${ \CC{4}{2}=6}$,~\eqref{initial_3} transforms into

\all{
& 6 \displaystyle  \CC{k+\ell-2}{\ell-1}   
+  
\displaystyle \CC{k+\ell+2}{k+1}
\nonumber \\
&=   
\displaystyle \sum_{\nu=0}^{3} \CC{\nu}{1}
\CC{k+\ell+1-\nu}{\ell-1}
+ 
\displaystyle \sum_{\nu=0}^{3} \CC{\nu}{1}
\CC{k+\ell+1-\nu}{k-1}
.}{initial_4}
The right hand side has six nonzero terms: 
\al{
& \displaystyle \sum_{\nu=0}^{3} \CC{\nu}{1}
\CC{k+\ell+1-\nu}{\ell-1}
+ 
\displaystyle \sum_{\nu=0}^{3} \CC{\nu}{1}
\CC{k+\ell+1-\nu}{k-1} 
\nonumber \\ 
&=  \displaystyle  \CC{k+\ell}{\ell-1}+2 \displaystyle  \CC{k+\ell-1}{\ell-1}+ 3\displaystyle  \CC{k+\ell-2}{\ell-1}
\nonumber \\
&+  \displaystyle  \CC{k+\ell}{k-1}+2 \displaystyle  \CC{k+\ell-1}{k-1}+ 3\displaystyle  \CC{k+\ell-2}{k-1}
}
Note that the third and sixth terms on the right hand side are identical. Replacing the right hand side of~\eqref{initial_4}  by this expression, we obtain
\all{
& 6 \displaystyle  \CC{k+\ell-2}{\ell-1}   
+  
\displaystyle \CC{k+\ell+2}{k+1}
\nonumber \\
&=   
 \displaystyle  \CC{k+\ell}{\ell-1}+2 \displaystyle  \CC{k+\ell-1}{\ell-1}+ 6\displaystyle  \CC{k+\ell-2}{\ell-1}
\nonumber \\
&+  \displaystyle  \CC{k+\ell}{k-1}+2 \displaystyle  \CC{k+\ell-1}{k-1}
.}{initial_5}
The first term on the left hand side cancels out with the third term on the right hand side and we get
\all{
& \displaystyle \CC{k+\ell+2}{k+1} =   
 \displaystyle  \CC{k+\ell}{\ell-1}+2 \displaystyle  \CC{k+\ell-1}{\ell-1}
+  \displaystyle  \CC{k+\ell}{k-1}+2 \displaystyle  \CC{k+\ell-1}{k-1}
.}{initial_6}
Dividing this equation by $(k+\ell-1)!$ and multiplying by ${(k+1)! (\ell+1)!}$, this transforms into
\all{
& (k+\ell+2)(k+\ell+1)(k+\ell) =   
 (k+\ell)\ell (\ell+1)
\nonumber \\
&+ 2 \ell (\ell+1) (k+1)+ (k+\ell) k (k+1)+ 2 k(k+1)(\ell+1)
.}{initial_7}
The second and fourth terms on the right hand side have the factor ${2(k+1)(\ell+1)}$ in common, and can be summed up and re-written as a single term~${2(k+1)(\ell+1)(k+\ell)}$. Then every term will share the factor~${(k+\ell)}$. dividing all terms by this factor, we obtain
\all{
& (k+\ell+2)(k+\ell+1)  \nonumber \\
&=   
 \ell (\ell+1)+   k (k+1)+ 2  (k+1)(\ell+1) 
.}{initial_8}
The right hand side can be simplified through the following successive steps
\all{ 
& \ell (\ell+1)+   k (k+1)+ 2  (k+1)(\ell+1) 
\nonumber \\
&= \ell (\ell+1)+    (k+1)(\ell+1) + k (k+1)+(k+1)(\ell+1) 
\nonumber \\
&= (\ell+1)(k+\ell+1)+ (k+1)(k+\ell+1) 
\nonumber \\
&= (k+\ell+2)(k+\ell+1)
.}{initial_9}
This means that~\eqref{initial_8} holds, so~\eqref{initial_1} is true, which means that the initial step of the induction is valid.

We now have to prove the inductive step. We return to~\eqref{objective_sum_1} which is repeated here for convenience of  reference:
\all{
&\displaystyle \frac{ (k+2) \CC{k+\ell+2}{\ell} }{ (\ell+1)}
-\displaystyle  \CC{2\beta+2}{\beta+1} 
 \CC{k+\ell-2\beta}{\ell-\beta}  \nonumber \\
&= 
\sum_{\mu=\beta+1}^k \CC{\mu+\beta+1}{\beta}
\CC{k+\ell-\beta-\mu}{\ell-\beta} 
\nonumber \\
&+ \sum_{\mu=\beta+1}^\ell \CC{\mu+\beta+1}{\beta}
\CC{k+\ell-\beta-\mu}{k-\beta} 
.}{objective_sum_3}
This can be equivalently expressed as follows:
\all{
&\displaystyle \frac{ (k+2) \CC{k+\ell+2}{\ell} }{ (\ell+1)}
=\displaystyle  \CC{2\beta+2}{\beta+1} 
 \CC{k+\ell-2\beta}{\ell-\beta}  \nonumber \\
&+ 
\sum_{\mu=\beta+1}^k \CC{\mu+\beta+1}{\beta}
\CC{k+\ell-\beta-\mu}{\ell-\beta} 
\nonumber \\
&+ \sum_{\mu=\beta+1}^\ell \CC{\mu+\beta+1}{\beta}
\CC{k+\ell-\beta-\mu}{k-\beta} 
.}{objective_sum_4}
We assume that this is true for $\beta$, and we prove the validity of the following, which is the same equality for ${\beta+1}$:
\all{
&\displaystyle \frac{ (k+2) \CC{k+\ell+2}{\ell} }{ (\ell+1)}
=\displaystyle  \CC{2\beta+4}{\beta+2} 
 \CC{k+\ell-2\beta-2}{\ell-\beta-1}  \nonumber \\
&+ 
\sum_{\mu=\beta+2}^k \CC{\mu+\beta+2}{\beta+1}
\CC{k+\ell-\beta-\mu-1}{\ell-\beta-1} 
\nonumber \\
&+ \sum_{\mu=\beta+2}^\ell \CC{\mu+\beta+2}{\beta+1}
\CC{k+\ell-\beta-\mu-1}{k-\beta-1} 
.}{inductive_step_1}
Using~\eqref{objective_sum_4} and noting that the upper bounds of the sums can equivalently be written as infinity (since the binomial coefficients impose the bounds automatically by making terms beyond the upper bound vanish), this can be recast as follows
\all{
&
\displaystyle  \CC{2\beta+2}{\beta+1} 
 \CC{k+\ell-2\beta}{\ell-\beta}  \nonumber \\
&+ 
\sum_{\mu=\beta+1}^{\infty} \CC{\mu+\beta+1}{\beta}
\CC{k+\ell-\beta-\mu}{\ell-\beta} 
\nonumber \\
&+ \sum_{\mu=\beta+1}^{\infty} \CC{\mu+\beta+1}{\beta}
\CC{k+\ell-\beta-\mu}{k-\beta} 
=\displaystyle  \CC{2\beta+4}{\beta+2} 
 \CC{k+\ell-2\beta-2}{\ell-\beta-1}  \nonumber \\
&+ 
\sum_{\mu=\beta+2}^{\infty} \CC{\mu+\beta+2}{\beta+1}
\CC{k+\ell-\beta-\mu-1}{\ell-\beta-1} 
\nonumber \\
&+ \sum_{\mu=\beta+2}^{\infty}  \CC{\mu+\beta+2}{\beta+1}
\CC{k+\ell-\beta-\mu-1}{k-\beta-1} 
.}{inductive_step_2}

Note that with a change in the summation index, we have the following equivalent forms for the sums:
\all{
\begin{cases}
\displaystyle \sum_{\mu=\beta+1}^{\infty}  \! \! \! \! \! \CC{\mu+\beta+1}{\beta}
\CC{k+\ell-\beta-\mu}{\ell-\beta} &= 
\displaystyle \sum_{\nu=2 \beta+2}^{\infty}  \! \! \! \! \! \CC{\nu}{\beta}
\CC{k+\ell+1-\nu}{\ell-\beta} 
\\ \\
\displaystyle \sum_{\mu=\beta+1}^{\infty}  \! \! \! \! \! \CC{\mu+\beta+1}{\beta}
\CC{k+\ell-\beta-\mu}{k-\beta} &= 
\displaystyle \sum_{\nu=2 \beta+2}^{\infty}  \! \! \! \! \! \CC{\nu}{\beta}
\CC{k+\ell+1-\nu}{k -\beta} 
\\ \\
\displaystyle \sum_{\mu=\beta+2}^{\infty}   \! \! \! \! \!  \CC{\mu+\beta+2}{\beta+1}
\CC{k+\ell-\beta-\mu-1}{\ell-\beta-1} &= 
\displaystyle \sum_{\nu=2\beta+4}^{\infty}  \! \! \! \! \! \CC{\nu}{\beta+1}
\CC{k+\ell+1-\nu}{\ell-\beta-1}
\\ \\
\displaystyle \sum_{\mu=\beta+2}^{\infty}   \! \! \! \! \!   \CC{\mu+\beta+2}{\beta+1}
\CC{k+\ell-\beta-\mu-1}{k-\beta-1}& =
\displaystyle \sum_{\nu=2\beta+4}^{\infty}  \! \! \! \! \!  \CC{\nu}{\beta+1}
\CC{k+\ell+1-\nu}{k-\beta-1}
\end{cases}
}{list_binoms}

Let us rearrange the terms of~\eqref{inductive_step_2} and break the sums into two distinct parts and rewrite  that equation  as follows
\all{
&
\displaystyle  \CC{2\beta+2}{\beta+1} 
 \CC{k+\ell-2\beta}{\ell-\beta} - \displaystyle  \CC{2\beta+4}{\beta+2} 
 \CC{k+\ell-2\beta-2}{\ell-\beta-1} 
 \nonumber \\
&+ 
\displaystyle \sum_{\nu=0}^{\infty}  \CC{\nu}{\beta}
\CC{k+\ell+1-\nu}{\ell-\beta} 
-\displaystyle \sum_{\nu=0}^{2\beta+1}    \CC{\nu}{\beta}
\CC{k+\ell+1-\nu}{\ell-\beta} 
\nonumber \\
&+ \displaystyle \sum_{\nu=0}^{\infty}  \CC{\nu}{\beta}
\CC{k+\ell+1-\nu}{k -\beta}  -
\displaystyle \sum_{\nu=0}^{2\beta+1}    \CC{\nu}{\beta}
\CC{k+\ell+1-\nu}{k -\beta} 
\nonumber \\
&=  
\displaystyle \sum_{\nu=0}^{\infty}  \CC{\nu}{\beta+1}
\CC{k+\ell+1-\nu}{\ell-\beta-1}
- 
\displaystyle \sum_{\nu=0}^{2\beta+3}  \CC{\nu}{\beta+1}
\CC{k+\ell+1-\nu}{\ell-\beta-1}
\nonumber \\
&+ \displaystyle \sum_{\nu=0}^{\infty}     \CC{\nu}{\beta+1}
\CC{k+\ell+1-\nu}{k-\beta-1}
-  \displaystyle \sum_{\nu=0}^{2\beta+3}   \CC{\nu}{\beta+1}
\CC{k+\ell+1-\nu}{k-\beta-1}
.}{inductive_step_4}

Following the same steps that allowed us to  obtain~\eqref{s1_temp_5}  from~\eqref{s1_temp_1}, we obtain the following identities:
\all{
\begin{cases}
\displaystyle \sum_{\nu=0}^{\infty}  \CC{\nu}{\beta}
\CC{k+\ell+1-\nu}{\ell-\beta}  = \CC{k+\ell+2}{\ell+1}
\\ \\
\displaystyle \sum_{\nu=0}^{\infty}  \CC{\nu}{\beta}
\CC{k+\ell+1-\nu}{k -\beta}=  \CC{k+\ell+2}{k+1}
\\ \\
\displaystyle \sum_{\nu=0}^{\infty}  \CC{\nu}{\beta+1}
\CC{k+\ell+1-\nu}{\ell-\beta-1} =  \CC{k+\ell+2}{\ell+1}
\\ \\
 \displaystyle \sum_{\nu=0}^{\infty}     \CC{\nu}{\beta+1}
\CC{k+\ell+1-\nu}{k-\beta-1} =  \CC{k+\ell+2}{k+1}
\end{cases}
}{list_binoms_identity}
It is obvious that the right hand sides are  all identical. So all the sums in~\eqref{inductive_step_4} whose upper bound is infinity cancel out, and the inductive step reduces to proving the validity of the following 
\all{
&
\displaystyle  \CC{2\beta+2}{\beta+1} 
 \CC{k+\ell-2\beta}{\ell-\beta} - \displaystyle  \CC{2\beta+4}{\beta+2} 
 \CC{k+\ell-2\beta-2}{\ell-\beta-1} 
 \nonumber \\
&
-\displaystyle \sum_{\nu=0}^{2\beta+1}    \CC{\nu}{\beta}
\CC{k+\ell+1-\nu}{\ell-\beta}  -
\displaystyle \sum_{\nu=0}^{2\beta+1}    \CC{\nu}{\beta}
\CC{k+\ell+1-\nu}{k -\beta} 
\nonumber \\
&=  
- 
\displaystyle \sum_{\nu=0}^{2\beta+3}  \CC{\nu}{\beta+1}
\CC{k+\ell+1-\nu}{\ell-\beta-1}
 -  \displaystyle \sum_{\nu=0}^{2\beta+3}   \CC{\nu}{\beta+1}
\CC{k+\ell+1-\nu}{k-\beta-1}
.}{inductive_step_5}

%

First let us focus on the last term of every sum appearing in this equation. The last term of the first sum on the  left  hand side is $\CC{2\beta+1}{\beta} \CC{k+\ell-2\beta}{\ell-\beta}$. The last term of the second  sum on the  left hand side is $\CC{2\beta+1}{\beta} \CC{k+\ell-2\beta}{k-\beta}$, which is the same as $\CC{2\beta+1}{\beta} \CC{k+\ell-2\beta}{\ell-\beta}$ by the basic properties of the binomial coefficient . The last term of the first sum on the right hand side is $\CC{2\beta+3}{\beta+1} \CC{k+\ell-2\beta-2}{\ell-\beta-1}$. The last term of the second sum on the right hand side is $\CC{2\beta+3}{\beta+1} \CC{k+\ell-2\beta-2}{k-\beta-1}$, which is the same as $\CC{2\beta+3}{\beta+1} \CC{k+\ell-2\beta-2}{\ell-\beta-1}$. Separating the last terms, we can write~\eqref{inductive_step_5}  in the following form:

\all{
&
\displaystyle  \CC{2\beta+2}{\beta+1} 
 \CC{k+\ell-2\beta}{\ell-\beta} - \displaystyle  \CC{2\beta+4}{\beta+2} 
 \CC{k+\ell-2\beta-2}{\ell-\beta-1} 
 \nonumber \\
&
-2 \CC{2\beta+1}{\beta} \CC{k+\ell-2\beta}{\ell-\beta}
\nonumber \\ &
-\displaystyle \sum_{\nu=0}^{2\beta }    \CC{\nu}{\beta}
\CC{k+\ell+1-\nu}{\ell-\beta}  -
\displaystyle \sum_{\nu=0}^{2\beta }    \CC{\nu}{\beta}
\CC{k+\ell+1-\nu}{k -\beta} 
\nonumber \\
&=  
-2 \CC{2\beta+3}{\beta+1} \CC{k+\ell-2\beta-2}{\ell-\beta-1} 
\nonumber \\
&
- 
\displaystyle \sum_{\nu=0}^{2\beta+2}  \CC{\nu}{\beta+1}
\CC{k+\ell+1-\nu}{\ell-\beta-1}
 -  \displaystyle \sum_{\nu=0}^{2\beta+2}   \CC{\nu}{\beta+1}
\CC{k+\ell+1-\nu}{k-\beta-1}
.}{inductive_step_6}

Now let us define: 
\all{
\begin{cases}
\Pi_1  \stackrel{\text{def}}{=}   &\displaystyle  \CC{2\beta+2}{\beta+1} 
 \CC{k+\ell-2\beta}{\ell-\beta} 
-2 \CC{2\beta+1}{\beta} \CC{k+\ell-2\beta}{\ell-\beta}   \\ \\
 \Pi_2  \stackrel{\text{def}}{=} &
 \displaystyle  \CC{2\beta+4}{\beta+2} 
 \CC{k+\ell-2\beta-2}{\ell-\beta-1} 
-2 \CC{2\beta+3}{\beta+1} \CC{k+\ell-2\beta-2}{\ell-\beta-1}  .
 \end{cases}
}{define_pis}

We can rewrite~\eqref{inductive_step_6} in the following compact form:

\all{
&
\Pi_2-\Pi_1
+\displaystyle \sum_{\nu=0}^{2\beta }    \CC{\nu}{\beta}
\CC{k+\ell+1-\nu}{\ell-\beta}  +
\displaystyle \sum_{\nu=0}^{2\beta }    \CC{\nu}{\beta}
\CC{k+\ell+1-\nu}{k -\beta} 
\nonumber \\
&=   
\displaystyle \sum_{\nu=0}^{2\beta+2}  \CC{\nu}{\beta+1}
\CC{k+\ell+1-\nu}{\ell-\beta-1}
 +  \displaystyle \sum_{\nu=0}^{2\beta+2}   \CC{\nu}{\beta+1}
\CC{k+\ell+1-\nu}{k-\beta-1}
.}{inductive_step_7}

We proceed by first proving that both $\Pi_1$ and $\Pi_2$ are identical to zero. For $\Pi_1$ we need to show that the following holds: 
\eq{
\CC{2\beta+2}{\beta+1}=2\CC{2\beta+1}{\beta}
.}
We begin from the left hand side:
\al{
\CC{2\beta+2}{\beta+1} & = \displaystyle \frac{(2\beta+2)!}{(\beta+1)! (\beta+1)!}
= (2\beta+2) \displaystyle \frac{(2\beta+1)!}{(\beta+1)! (\beta+1)!} \nonumber \\
& = 
2 \displaystyle \frac{(2\beta+1)!}{\beta ! (\beta+1)!}= 2\CC{2\beta+1}{\beta}.
}
To prove that $\Pi_2$ is zero, we must prove
\eq{
 \displaystyle  \CC{2\beta+4}{\beta+2} 
 =2 \CC{2\beta+3}{\beta+1}  
.}
Starting  from the left hand side, we get:
\al{
\CC{2\beta+4}{\beta+2} & = \displaystyle \frac{(2\beta+4)!}{(\beta+2)! (\beta+2)!}
= (2\beta+4) \displaystyle \frac{(2\beta+3)!}{(\beta+2)! (\beta+2)!} \nonumber \\
& = 
2 \displaystyle \frac{(2\beta+3)!}{(\beta+2)! (\beta+1)!}= 2 \CC{2\beta+3}{\beta+1} .
}

Setting $\Pi_1=\Pi_2=0$ in~\eqref{inductive_step_7}, the inductive step reduces to proving that the following is true:

\all{
&
 \displaystyle \sum_{\nu=0}^{2\beta }    \CC{\nu}{\beta} \Bigg[ 
\CC{k+\ell+1-\nu}{\ell-\beta}  +
\CC{k+\ell+1-\nu}{k -\beta}  \Bigg] 
\nonumber \\
&=   
\displaystyle \sum_{\nu=0}^{2\beta+2}  \CC{\nu}{\beta+1} \Bigg[ 
\CC{k+\ell+1-\nu}{\ell-\beta-1}
 + 
\CC{k+\ell+1-\nu}{k-\beta-1} \Bigg] 
.}{inductive_step_8}
Let us define: 
\al{
f(\beta,k,\ell) \stackrel{\text{def}}{=} 
 \displaystyle \sum_{\nu=0}^{2\beta }    \CC{\nu}{\beta} \Bigg[ 
\CC{k+\ell+1-\nu}{\ell-\beta}  +
\CC{k+\ell+1-\nu}{k -\beta}  \Bigg] 
.}
If we prove that $f(\beta,k,\ell)$ does not depend on~$\beta$, then the validity of~\eqref{inductive_step_8} immediately follows. 

We define the following generating function  
\al{
g(\beta,k,x)  \stackrel{\text{def}}{=}  \displaystyle \sum_{\ell=0 }^{\infty} f(\beta,k,\ell) x^{\ell}.
}
Then the coefficient of $x^{\ell}$  in this series for each specific value of~$\ell$ will be equal to the value of the function 
 $f(\cdot)$. Note that to prove that $f(\beta,k,\ell)$ is independent of $\beta$, it suffices to show that $g(\beta,k,x)$ is independent of $\beta$, which is what we intend to ascertain.

Let us  calculate $g(\beta,k,x)$ explicitly. We have:
\al{
g(\beta,k,x)  &= \displaystyle  \displaystyle \sum_{\ell=0 }^{\infty} 
 \displaystyle \sum_{\nu=0}^{2\beta }    \CC{\nu}{\beta} \Bigg[ 
\CC{k+\ell+1-\nu}{\ell-\beta}  +
\CC{k+\ell+1-\nu}{k -\beta}  \Bigg]   x^{\ell} .
}
Interchanging the order of summations, and noting that the sum on $\nu$ starts having nonzero terms from ${\nu=\beta}$, we can rewrite ${ g(\beta,k,x)}$ in the following form:
\al{
& g(\beta,k,x)  = \displaystyle \sum_{\nu=\beta}^{2\beta }  \CC{\nu}{\beta}
  \displaystyle \sum_{\ell=0 }^{\infty} 
  \Bigg[ 
\CC{k+\ell+1-\nu}{\ell-\beta}  +
\CC{k+\ell+1-\nu}{k -\beta}  \Bigg]   x^{\ell}  \nonumber \\
&=  \displaystyle \sum_{\nu=\beta}^{2\beta }  \CC{\nu}{\beta}
\Bigg[   x^{-k-1+\nu}   \displaystyle \sum_{\ell=0 }^{\infty} 
 \CC{k+\ell+1-\nu}{k+\beta+1-\nu}  x^{k+\ell+1-\nu}  \nonumber \\ &
+ x^{-k-1+\nu}   \displaystyle \sum_{\ell=0 }^{\infty}  \CC{k+\ell+1-\nu}{k -\beta}    x^{k+\ell+1-\nu}    \Bigg] .
}
Denoting $k+\ell+1-\nu$ by $\theta$ and using~\eqref{taylor_1}, we obtain
\all{
& g(\beta,k,x)  =   \displaystyle \sum_{\nu=\beta}^{2\beta }  \CC{\nu}{\beta}
\Bigg[   x^{-k-1+\nu}   \displaystyle \sum_{\theta }
 \CC{\theta}{k+\beta+1-\nu}  x^{\theta}  \nonumber \\ &
+ x^{-k-1+\nu}   \displaystyle \sum_{\ell=0 }^{\infty}  \CC{\theta}{k -\beta}    x^{\theta}    \Bigg]  
\nonumber \\ &
=   \displaystyle \sum_{\nu=\beta}^{2\beta }  \CC{\nu}{\beta}  x^{-k-1+\nu} 
\Bigg[ \displaystyle \frac{x^{k+\beta+1-\nu}}{(1-x)^{k+\beta+2-\nu}} 
+ \displaystyle      \frac{x^{k-\beta}}{(1-x)^{k+1-\beta}}      \Bigg] 
\nonumber \\ &
=   \displaystyle \sum_{\nu=\beta}^{2\beta }  \CC{\nu}{\beta}    
\Bigg[ \displaystyle \frac{x^{ \beta}}{(1-x)^{k+\beta+2-\nu}} 
+ \displaystyle      \frac{x^{\nu-1-\beta}}{(1-x)^{k+1-\beta}}      \Bigg] 
\nonumber \\ &
=  \displaystyle \frac{x^{-1}}{(1-x)^{k+2}} \nonumber \\ &
\times \displaystyle \sum_{\nu=\beta}^{2\beta }  \CC{\nu}{\beta}
\Bigg[ x^{\beta+1} (1-x)^{\nu-\beta} + (1-x)^{\beta+1} x^{\nu-\beta}
   \Bigg] 
}{inductive_step_10}
Now let us define
\eq{
h(\beta,x) \stackrel{\text{def}}{=} 
 \displaystyle \sum_{\nu=\beta}^{2\beta }  \CC{\nu}{\beta}
\Bigg[ x^{\beta+1} (1-x)^{\nu-\beta} + (1-x)^{\beta+1} x^{\nu-\beta}
   \Bigg] 
.}
Then~\eqref{inductive_step_10} can be written as follows: 
\al{
g(\beta,k,x)= \displaystyle \frac{x^{-1}}{(1-x)^{k+2}}  h(\beta,x)
.}
Now we take a step further and show that not only $h(\beta,x)$ is independent of $\beta$, but it is also independent of $x$. In fact, we are going to prove the following algebraic identity for any $x$ except zero and unity: 
\all{
 \displaystyle \sum_{\nu=\beta}^{2\beta }  \CC{\nu}{\beta}
\Bigg[ x^{\beta+1} (1-x)^{\nu-\beta} + (1-x)^{\beta+1} x^{\nu-\beta}
   \Bigg] =1
.}{new_objective}
Note that once this is proven, then $g(\beta,k,x)$ will be independent of $\beta$, and so will $f(\beta,k,\ell)$, whose  immediate consequence is the validity of~\eqref{inductive_step_8}, which would conclude the proof.

To show that~\eqref{new_objective} is true, we are going to prove that the coefficient of $x^0$ on the left hand side is equal to unity, and the coefficients of all other powers of $x$ are identical to zero.

 For $x^0$, observe that the first term in the summand cannot yield an $x^0$ term because the powers of $x$ will be at least ${\beta+1}$. The second term can only yield an $x^0$ term if $\nu=\beta$, and the coefficient of $x^0$ for the whole sum will be the coefficient of $x^0$ in the factor ${(1-x)^{\beta+1}}$ times the factor $\CC{\beta}{\beta}$, which are both equal to unity. We conclude that the coefficient of $x^0$ on the left hand side is equal to that of the right hand side. 

Now let us focus on the coefficient of $x^k$ for general ${k \neq 0}$. Note that the summands yield $x^k$ terms for ${1 \leq k \leq 2\beta+1}$. 
We divide this range  into two parts. First we show that the coefficient of $x^k$ is zero for $1 \leq k \leq \beta$, and then we will show it is zero for~${\beta+1 \leq k \leq 2\beta+1}$. 

Only the second  summand can yield $x^k$ terms with~$1 \leq k \leq \beta$. For each $\nu$, the coefficient of $x^k$ will be equal to $\CC{\nu}{\beta}$ times the coefficient of $x^{k-(\nu-\beta)}$ that $(1-x)^{\beta+1}$ yields, which is equal to~${\CC{\beta+1}{k+\beta-\nu}(-1)^{k+\beta+\nu}}$. So the total coefficient of $x^k$ for~$1 \leq k \leq \beta$ equals
\al{
(-1)^{k+\beta} \sum_{\nu=\beta}^{2\beta} \CC{\nu}{\beta} \CC{\beta+1}{k+\beta-\nu} (-1)^{\nu}
.}
In terms of $\alpha=\nu-\beta$, this takes the following  simpler form
\al{
(-1)^{k} \sum_{\alpha=0}^{\beta} \CC{\alpha+\beta}{\beta}  \CC{\beta+1}{k-\alpha} (-1)^{\alpha}
.}
Since $k\leq \beta$, we can use $k$ is the summation limit instead of $\beta$. Also the lower bound for $\alpha$ can be set to $-\infty$ instead of zero, since the binomial coefficient~$ \CC{\alpha+\beta}{\beta}$ automatically rules out the terms for  negative values of $\alpha$. So the coefficient of $x^k$ is equal to
\al{
(-1)^{k} \sum_{\alpha=0}^{k} \CC{\alpha+\beta}{\beta}  \CC{\beta+1}{k-\alpha} (-1)^{\alpha}
.}
Following the same steps that allowed us to  obtain~\eqref{s1_temp_5}  from~\eqref{s1_temp_1}, we know that the sum is equal to the coefficient of $x^k$ in the power series which is the product of the following two power series. The first one whose coefficient of $x^{\alpha}$ is $\CC{\alpha+\beta}{\beta} $, and the second one whose coefficient of $x^{\alpha}$ is  $ \CC{\beta+1}{\alpha} (-1)^{\alpha}$. The first function  is thus $\frac{1}{(1-x)^{\beta+1}}$ and the second one is $(1-x)^{\beta+1}$. Their product is unity, whose coefficients of $x^k$ are all zero for~$1\leq k \leq \beta$. 

Now we focus on  the coefficients of $x^k$ for  ${\beta+1 \leq k \leq 2\beta+1}$.
 The coefficient of $x^k$ that the first term in the summand yields is equal to the coefficient of $x^{k-(\beta+1)}$ in the binomial expansion of~${(1-x)^{\nu-\beta}}$, which is equal to the following:
\eq{
\CC{\nu-\beta}{k-\beta-1} (-1)^{k-\beta-1}.
}
The coefficient of $x^k$ from the second term of the summand is obtained similarly, and is equal to the following:
\eq{
\CC{\beta+1}{k+\beta-\nu} (-1)^{k+\beta-\nu}.
}
Putting these two together, we get the overall coefficient of $x^k$. We intended to show that it is equal to zero. Taking out a factor of $(-1)^{k-\beta}$,  the objective is now to prove the following: 
\all{
 \displaystyle \sum_{\nu=\beta}^{2\beta }  \CC{\nu}{\beta} 
\Bigg[ 
\CC{\nu-\beta}{k-\beta-1}- (-1)^{\nu} \CC{\beta+1}{k+\beta-\nu}
\Bigg] =0
.}{new_objective_2}

This can be equivalently expressed in the following form:

\all{
 \displaystyle \sum_{\nu=\beta}^{2\beta }  \CC{\nu}{\beta} 
\CC{\nu-\beta}{k-\beta-1}  =
 \displaystyle \sum_{\nu=\beta}^{2\beta }  (-1)^{\nu} \CC{\nu}{\beta} 
 \CC{\beta+1}{k+\beta-\nu} 
.}{new_objective_3}

We prove this equality by explicitly calculating both sides and showing that they are equal to the same thing. For the left hand side, first note that the following holds:
\al{
\CC{\nu}{\beta} 
\CC{\nu-\beta}{k-\beta-1}  =
\CC{k-1}{\beta} 
\CC{\nu}{k-1} 
.}
Using this, we have:
\all{
&  \displaystyle \sum_{\nu=\beta}^{2\beta }  \CC{\nu}{\beta} 
\CC{\nu-\beta}{k-\beta-1}  =
 \CC{k-1}{\beta}  \displaystyle \sum_{\nu=\beta}^{2\beta } \CC{\nu}{k-1} 
.}{left_1}
To prove this equality holds, first note that the following is true: 
\al{
 \CC{\nu}{k-1}=  \CC{\nu+1}{k} -  \CC{\nu}{k}
.}
Then the sum telescopes and we find the result as follows: 
\al{
 \CC{k-1}{\beta}  \displaystyle \sum_{\nu=\beta}^{2\beta } \CC{\nu}{k-1} & = 
 \CC{k-1}{\beta}  \displaystyle \sum_{\nu=\beta}^{2\beta } \Bigg[   \CC{\nu+1}{k} -  \CC{\nu}{k}  \Bigg]  
\nonumber \\
 &=  \CC{k-1}{\beta} \bigg[   \CC{2\beta+1}{k} -  \CC{\beta}{k}  \bigg] .
}
The second term is $ \CC{k-1}{\beta}   \CC{\beta}{k} $. This is identical to zero, because in order for the first factor to be nonzero,   $k$ must be greater than $\beta$, which makes the second factor vanish. So we obtain the following  result: 
\al{
 \displaystyle \sum_{\nu=\beta}^{2\beta }  \CC{\nu}{\beta} 
\CC{\nu-\beta}{k-\beta-1}  =
 \CC{k-1}{\beta}  \displaystyle \sum_{\nu=\beta}^{2\beta } \CC{\nu}{k-1} 
=  \CC{k-1}{\beta}     \CC{2\beta+1}{k} 
.}

The last step is to show that the right hand side of~\eqref{new_objective_3} is equal to the same thing. So we need to prove the validity of the following identity:
\all{
 \displaystyle \sum_{\nu=\beta}^{2\beta }  (-1)^{\nu} \CC{\nu}{\beta} 
 \CC{\beta+1}{k+\beta-\nu} 
= 
 \CC{k-1}{\beta}    \CC{2\beta+1}{k} .
}{final_Step}

To do so, we first need a generalization for the binomial coefficients to include negative arguments for the upper index. We denote the generalized version of the binomial coefficient  by $\CCC{x}{y}$, where it is defined in the following general form: 
\all{
\CCC{x}{y}=  \displaystyle  \frac{x (x-1) (x-2) \ldots [x-(y-1)]}{y!}
.
}{CCC_def}
Using this notation, one can easily verify that the following is true: 
\al{
\CC{\nu}{\beta}= (-1)^{\nu+\beta} \CCC{-\beta-1}{\nu-\beta}
.}
Replacing the factor $\CC{\nu}{\beta}$ in the summand on the  left  hand side of~\eqref{final_Step}, and denoting ${\nu-\beta}$ by $\eta$, we arrive at the following equivalent form for the identity  we seek to prove:
\all{
 \displaystyle \sum_{\eta=0}^{\beta }   \CCC{-\beta-1}{\eta}   \CC{\beta+1}{k  -\eta}
= 
(-1)^{\beta}  \CC{k-1}{\beta}    \CC{2\beta+1}{k} .
}{fin_temp_1}
In~\cite{god_bless_andersen}, the following identity is proved through induction: 
\all{
 \displaystyle \sum_{\eta=0}^{w }   \CCC{-x}{y} 
 \CCC{x }{z-y } 
=  \displaystyle \frac{z-w}{z}
 \CCC{-x-1}{w}    \CCC{x}{z-w} .
}{andersen_identity}
Comparing this with~\eqref{fin_temp_1}, we find that the following is true
\all{
 \displaystyle \sum_{\eta=0}^{\beta }   \CCC{-\beta-1}{\eta}   \CC{\beta+1}{k  -\eta}
= 
\displaystyle \frac{k-\beta}{k}   \CCC{-\beta-2}{\beta}    \CC{\beta+1}{k-\beta} .
}{fin_temp_2}
Using~\eqref{CCC_def}, it is readily verifiable that the following holds: 
\all{
\CCC{-x}{y}=(-1)^{y} \CCC{x+y-1}{y}
.}{CCC_prop}
Using this, we obtain
\al{
  \CCC{-\beta-2}{\beta} = (-1)^{\beta} \CC{2\beta+1}{\beta}
.}
Replacing the factor $  \CCC{-\beta-2}{\beta}$ in~\eqref{fin_temp_2}, we arrive at
\all{
 \displaystyle \sum_{\eta=0}^{\beta }   \CCC{-\beta-1}{\eta}   \CC{\beta+1}{k  -\eta}
= (-1)^{\beta}
\displaystyle \frac{k-\beta}{k}   \CC{2\beta+1}{\beta}     \CC{\beta+1}{k-\beta} .
}{fin_temp_3}
Through the following algebraic steps, we can recast the right hand side:
\al{
&(-1)^{\beta} \displaystyle  \frac{k-\beta}{k}  \CC{2\beta+1}{\beta}     \CC{\beta+1}{k-\beta}  
\nonumber \\ 
&= (-1)^{\beta} \displaystyle  \frac{k-\beta}{k}
\displaystyle  \frac{(2\beta+1)!}{\beta! {(\beta+1)!}}
\displaystyle  \frac{ {(\beta+1)!}}{(k-\beta)! (2\beta+1-k)!}
\nonumber \\
&= (-1)^{\beta} \displaystyle  \frac{k-\beta}{k}  \displaystyle  \frac{1}{\beta!  (k-\beta)! } \displaystyle  \frac{k!}{k!}
\displaystyle  \frac{  (2\beta+1)! }{ (2\beta+1-k)!}
\nonumber \\
&=  (-1)^{\beta}  \displaystyle  \frac{(k-1)!}{\beta!  (k-\beta-1)! } \displaystyle  
\displaystyle  \CC{2\beta+1}{k}
=  (-1)^{\beta}  \displaystyle  \CC{k-1}{\beta}
\displaystyle  \CC{2\beta+1}{k}.
}
This is identical to the right hand side of~\eqref{fin_temp_1}. Hence the proof is complete.


\section{Degree Distribution of Shifted Linear Preferential Attachment} \label{app:nk_shifted}

The rate equation for the evolution of~$N_k$ is
\al{
&N_k(t+1)-N_k(t)=\displaystyle \frac{\beta(k-1+\theta)N_{k-1}}{2L(0)+2\beta t+ \theta N(t)} \nonumber \\
&- \displaystyle \frac{\beta(k+\theta)N_{k}}{2L(0)+2\beta t+ \theta N(t)}
+ \delta_{k,\beta}
.}
Replacing~$N_k$ by~$N n_k$, in the steady state we have
\al{
&n_k=\displaystyle \frac{\beta(k-1+\theta)n_{k-1}}{2\beta + \theta }  - \displaystyle \frac{\beta(k+\theta)n_{k}}{2\beta  + \theta  }
+ \delta_{k,\beta}
.}
This can be rearranged and transformed into
\al{
&n_k \Big[ 2+k+\theta(1+\frac{1}{\beta})\Big] =\displaystyle   (k-1+\theta)n_{k-1} + \Big[ 2+ \frac{\theta}{\beta} \Big] \delta_{k,\beta}
,}
or equivalently, 
\al{
&n_k   =\displaystyle \frac{   (k-1+\theta)}{ 2+k+\theta(1+\frac{1}{\beta})}n_{k-1} +  \displaystyle \frac{ 2+ \frac{\theta}{\beta} }{ 2+\beta+\theta(1+\frac{1}{\beta})} \delta_{k,\beta}
.}
Solving this recursion relation is straightforward. The first term is
\al{
&n_\beta   = \displaystyle \frac{ 2+ \frac{\theta}{\beta} }{ 2+\beta+\theta(1+\frac{1}{\beta})} 
.}
And for~$k>\beta$, we have
\al{
n_k =  n_\beta \prod_{\mu=\beta}^{k-1} \frac{ \mu+\theta}{\mu+3+\theta(1+\frac{1}{\beta})}
.}
By the use of Gamma functions, this expression can be revamped into the following:
\al{
n_k =  n_\beta  \displaystyle \frac{\Gamma(k+\theta)}{\Gamma(\beta+\theta)}
\displaystyle \frac{\Gamma\left[ \beta+3+\theta(1+\frac{1}{\beta})\right] }{\Gamma\left[ k+3+\theta(1+\frac{1}{\beta})\right]}
.}
Denoting~${2+ \frac{\theta}{\beta} }$ by~$\nu$, we arrive at 
\al{
n_k =  \displaystyle \frac{\nu}{\beta+\theta+\nu}  \displaystyle \frac{\Gamma(k+\theta)}{\Gamma(\beta+\theta)}
\displaystyle \frac{\Gamma (\beta+1+\theta+\nu ) }{\Gamma ( k+1+\theta+\nu )}
.}
Using the properties ofthe Gamma function, this can be recast as follows 
\al{
n_k =  \displaystyle  \nu   \displaystyle \frac{\Gamma(k+\theta)}{\Gamma(\beta+\theta)}
\displaystyle \frac{\Gamma (\beta+\theta+\nu ) }{\Gamma ( k+1+\theta+\nu )}
.}


\section{Solving the Difference Equation~\eqref{nkl_temp_shifted_2_text}} \label{app:shifted_diff_solve}

\all{
& n_{k{\ell}} =  
\frac{(k-1+\theta)  n_{k-1,{\ell}} }{k+\ell+2\theta+\nu}  +  \frac{(\ell-1+\theta) n_{k,{\ell}-1} }{k+\ell+2\theta+\nu} \nonumber \\
&+ \left(\displaystyle \frac{\nu}{\beta+\theta+\nu}\right)   
\left( \displaystyle \frac{\Gamma (\beta+1+\theta+\nu )}{\Gamma(\beta+\theta)} \right)
\nonumber \\
& \times \frac{1 }{(k+\beta+2\theta+\nu)}  
\displaystyle \frac{\Gamma(k+\theta) }{\Gamma ( k+\theta+\nu )}
\delta(\ell,\beta)    
.}{nkl_temp_shifted_2}

Now define
\all{
m_{k{\ell}} \stackrel{\text{def}}{=} 
\frac{\Gamma(k+\ell+2\theta+\nu+1)}
{\Gamma(k+\theta)  \Gamma(\ell+\theta)} 
n_{k{\ell}}
.}{mkl_def_shifted}
The following relations are true by definition
\al{
\begin{cases}
\displaystyle  n_{k,{\ell}}
= \displaystyle  \frac{\Gamma(k+\theta)  \Gamma(\ell+\theta)} 
{\Gamma(k+\ell+2\theta+\nu+1)} m_{k,{\ell}}                        \\ \\
\displaystyle (k-1+\theta) n_{k-1,{\ell}}
= \displaystyle  \frac{\Gamma(k+\theta)  \Gamma(\ell+\theta)} 
{\Gamma(k+\ell+2\theta+\nu)}    m_{k-1,{\ell}}
                                   \\ \\
\displaystyle (\ell-1+\theta) n_{k,{\ell}-1}
= \displaystyle  \frac{\Gamma(k+\theta)  \Gamma(\ell+\theta)} 
{\Gamma(k+\ell+2\theta+\nu)}  m_{k,{\ell}-1}
\end{cases}
}
Plugging these into~\eqref{nkl_temp_shifted_2} and multiplying through by 
the factor~${\frac{\Gamma(k+\ell+2\theta+\nu+1)}
{\Gamma(k+\theta)  \Gamma(\ell+\theta)}}$ and replacing all~$\ell$s with~$\beta$ in the last term (which alters nothing, because of multiplication by the delta function), we obtain
\all{
&m_{k{\ell}} =  
m_{k-1,{\ell}}   
+ m_{k,{\ell}-1} \nonumber \\
&+ \left(\displaystyle \frac{\nu}{\beta+\theta+\nu}\right)   
\left( \displaystyle \frac{\Gamma (\beta+1+\theta+\nu )}{\Gamma(\beta+\theta)} \right)
\nonumber \\
& \times \frac{1 }{(k+\beta+2\theta+\nu)}  
\displaystyle \frac{\Gamma(k+\beta+2\theta+\nu+1) }{\Gamma ( k+\theta+\nu ) \Gamma(\beta+\theta)}
\delta(\ell,\beta)       
.}{nkl_temp_shifted_3}
For brevity, let us define
\al{
A \stackrel{\text{def}}{=} 
 \displaystyle \frac{\nu}{\beta+\theta+\nu}   
  \displaystyle \frac{\Gamma (\beta+1+\theta+\nu )}{\big[\Gamma(\beta+\theta)\big]^2}  
  =    \nu    
  \displaystyle \frac{\Gamma (\beta+\theta+\nu )}{\big[\Gamma(\beta+\theta)\big]^2}
}
So we arrive at the following difference equation
\all{
&m_{k{\ell}} =  
m_{k-1,{\ell}}   
+ m_{k,{\ell}-1} \nonumber \\
&+ A    
\displaystyle \frac{\Gamma(k+\beta+2\theta+\nu) }{\Gamma ( k+\theta+\nu )  }
\delta(\ell,\beta)       u(k-\beta-1)
.}{mkl_diff_shifted}
To solve this partial difference equation, we use the two-dimensional Z-transform as we did for the non-shifted version above. All steps and also the integrals encountered are identical to those of the non-shifted case, and we arrive at
\al{
m_{k{\ell}} = A \sum_{\mu=\beta+1}^k \displaystyle \frac{\Gamma(k+\beta+2\theta+\nu) }{\Gamma ( k+\theta+\nu )  } 
\CC{k+\ell-\beta-\mu}{\ell-\beta}
.}
Plugging this into~\eqref{mkl_def_shifted} we get
\all{
n_{k \ell} &= \displaystyle A \frac{\Gamma(k+\theta)  \Gamma(\ell+\theta)} 
{\Gamma(k+\ell+2\theta+\nu+1)} \nonumber \\
& \times 
\sum_{\mu=\beta+1}^k \displaystyle \frac{\Gamma(\mu+\beta+2\theta+\nu) }{\Gamma ( \mu+\theta+\nu )  } 
\CC{k+\ell-\beta-\mu}{\ell-\beta}
.}{shifted_nkl_use_app}

\bibliographystyle{apsrev4-1}
\bibliography{myrefs}

\end{document}